\documentclass[11pt]{article}
\usepackage[margin=1.1in]{geometry}
\usepackage[round]{natbib}
\usepackage{amsmath, amssymb, enumitem, bm, bbm, multirow, hyperref, graphicx, authblk, tikz, amsthm, mathrsfs}
\usepackage{booktabs, longtable}
\usepackage[noabbrev]{cleveref}
\usepackage{thmtools,thm-restate}

\hypersetup{colorlinks, breaklinks,urlcolor= magenta, linkcolor= blue, citecolor = magenta}

\newcommand{\E}{\mathbb{E}}
\newcommand{\R}{\mathbb{R}}
\newcommand{\N}{\mathbb{N}}

\newcommand{\F}{\mathcal{F}}

\newcommand{\Q}{\mathbb{Q}}
\renewcommand{\P}{\mathbb{P}}
\newcommand{\T}{\mathcal{T}}
\newcommand{\V}{\text{Var}}

\newcommand{\cbrace}[2]{\left[\left. #1 \,\right|\, #2 \right]}
\newcommand{\indep}{\perp \!\!\! \perp}
\newcommand{\1}{\mathbbm{1}}
\newcommand{\m}{\text{min}}
\allowdisplaybreaks
\makeatletter
\renewcommand\AB@affilsepx{, \protect\Affilfont}
\makeatother

\title{Partial Identification of Causal Effects for Endogenous Continuous Treatments}

\author{Abhinandan Dalal}
\author{Eric J. Tchetgen Tchetgen}
\affil{University of Pennsylvania}

\date{}

\newtheorem{theorem}{Theorem}
\newtheorem{corollary}{Corollary}
\newtheorem{assumption}{Assumption}
\newtheorem{lemma}{Lemma}

\newtheorem{proposition}{Proposition}
\newtheorem{result}{Result}
\newtheorem{remark}{Remark}

\begin{document}

\maketitle

\begin{abstract}
No unmeasured confounding is a common assumption when reasoning about counterfactual outcomes, but such an assumption may not be plausible in observational studies. Sensitivity analysis is often employed to assess the robustness of causal conclusions to unmeasured confounding, but existing methods are predominantly designed for binary treatments. In this paper, we provide natural extensions of two extensively used sensitivity frameworks -- the Rosenbaum and Marginal sensitivity models -- to the setting of continuous exposures. Our generalization replaces scalar sensitivity parameters with sensitivity functions that vary with exposure level, enabling richer modeling and sharper identification bounds. We develop a unified pseudo-outcome regression formulation for bounding the counterfactual dose–response curve under both models, and propose 
corresponding nonparametric estimators which have second order bias. These estimators accommodate modern machine learning methods for obtaining nuisance parameter estimators, which are shown to achieve $L^2$-consistency, minimax rates of convergence under suitable conditions. Our resulting estimators of bounds for the counterfactual dose-response curve are shown to be consistent and asymptotic normal allowing for a user-specified bound on the degree of uncontrolled exposure endogeneity.
We also offer a geometric interpretation that relates the Rosenbaum and Marginal sensitivity model 
and guides their practical usage in global versus targeted sensitivity analysis. The methods are validated through simulations and a real-data application on the effect of second-hand smoke exposure on blood lead levels in children. 
\end{abstract}

\section{Introduction} \label{intro}
Causal inference in observational studies commonly proceeds under the assumption of no unmeasured confounding (NUC), also known as no endogeneity: treatment assignment is effectively randomized within strata defined by a sufficiently rich set of observed covariates. Although this assumption enables identification of causal parameters and effective inference through propensity score matching \cite{rosenbaum2010design}, inverse-probability weighting \cite{horvitz1952generalization} and doubly robust and double-machine learning estimators \cite{robins1994estimation, van2006targeted, chernozhukov2018double} — it is fundamentally untestable from the observed data without an alternative assumption 
and may be violated in practice.

Lack of sufficient control for confounding can lead to different conclusions from observational and experimental studies, and hence can threaten the validity of conclusions obtained from the former. For example, observational studies had raised concerns about the possible harm associated with the use of calcium channel blockers like nifedipine (\cite{psaty1995risk, pahor2000health}) with respect to myocardial infarction (commonly known as heart attack), only to be settled definitively about a decade later via randomized controlled trials that nifedipine was safe (\cite{psaty2004contemplating}). In-depth analysis revealed that calcium channel blockers were used to treat hypertension, which in itself was a risk factor associated with heart attacks (see \cite{rutter2007identifying} for more details).

To assess the robustness of causal conclusions about violations of NUC, researchers increasingly rely on sensitivity analysis to quantify the impact of \textit{potential} unmeasured confounding. In response to concerns about unmeasured confounding dating back to \cite{fisher1958cigarettes} about the impact of smoking on lung cancer, \cite{cornfield1959smoking} conducted the first known formal sensitivity analysis to establish that an unmeasured confounder would need to be nine times more prevalent among smokers than non-smokers to explain the observed association, a magnitude deemed implausible. Since then, there have been immense interest in developing methods for sensitivity analysis, the works of \cite{robins1999association, rosenbaum2010design, vanderweele2011bias, imbens2015causal, chernozhukov2021omitted, kallus2019interval, zhao2019sensitivity, bekerman2024planning, nabi2024semiparametric} are some notable works in this regard.

However, most existing works have primarily focused on binary exposures. Continuous exposures provide unique challenges to causal inference problems, since (i) the probability of observing any $\{T=t\}$ is zero, and (ii) there would be infinitely many potential outcome means $\{\E[Y(t)]\}_{t\in\T}$ to identify. Existing literature on causal inference with continuous exposure have been considered in \cite{lu2001matching, fogarty2021biased, zhang2023statistical} etc. for matching methods, and \cite{kennedy2017non, colangelo19052025, schindl2024incremental} etc. for double robust estimation, but these works have not included sensitivity analysis to NUC. CATE estimation with continuous covariates also deploys similar techniques as continuous exposures, as illustrated in \cite{kennedy2023towards} and \cite{semenova2021debiased}. 

Sensitivity analysis with continuous exposures in matched studies has recently been considered in \cite{zhang2024sensitivity} and \cite{frazier2024bias}, and interestingly, the first work establishes hardness results for sensitivity analysis in matched studies with continuous exposure and outcome. A more general class of $f$-sensitivity models have also been considered in \cite{jin2022sensitivity} and \cite{frauen2024neural}, where they bound a more general convex-link $f$ of  the odd's ratio, instead of the identity link, to accommodate the unbounded support of unmeasured confounders. \cite{bonvini2022sensitivity} also studied sensitivity analysis in marginal structural models with continuous exposures, introducing approaches based on propensity scores (see also \cite{jesson2022scalable, marmarelis2023partial}), outcome regression, and subset confounding that differ from standard formulations; notably, they bound density ratios in the former model rather than odds ratios (see Section~\ref{comparison} and Appendix \ref{bkvw} for further details), and leverage low-dimensional structure to achieve $\sqrt{n}$-rate inference. In this work, we deviate from their strategy and instead consider double-robust partial identification of the dose-response curve, and we directly estimate corresponding bounds under less restrictive non-parametric regression conditions. 



Although the need for sensitivity analysis is widely accepted, there remains a broad range of sensitivity models to choose from. We focus on continuous exposure extensions of two well-established canonical binary exposure models: Rosenbaum's model (\cite{rosenbaum2002observational}), and the marginal model of Tan (\cite{tan2006distributional}). Both models, originally devised for binary exposures, bound the odds ratio of receiving exposure for different levels of an unmeasured confounder $U$. Moreover, both models include the NUC data-distribution within permissible distributions for any non-trivial value of their sensitivity parameter. We extend these models to the continuous exposure case, and derive novel identification bounds for the dose-response curve under both models. Our work builds upon the existing extensions to multi-level exposures of \cite{rosenbaum1989sensitivity}, however, we develop our framework for partial inference to respect key inequality constraints anchoring our sensitivity analyses. It may be worthwhile to mention that both aforementioned models belong to a class of $L^\infty$ models, which bounds the maximum deviation of the data-distribution from NUC. Recent works have also explored $L^2$ class of models, which arguably being less interpretable than $L^\infty$ models (\cite{zhang2024linftyl2sensitivityanalysiscausal, huang2025variance}), produce narrower bounds for the potential outcome mean, and could hence be appealing to researchers. Nevertheless, in this work, we focus on the specified $L^\infty$ models in the case for continuous exposures and defer the development of analogous bounds under $L^2$ or more general models to future work.

\subsection{Our Contributions}

We delineate our contributions in this paper as the following:
\begin{enumerate}[label = (\roman*)]
    \item We introduce the notion of a \textit{sensitivity function} for continuous exposures, allowing bounds to vary with exposure levels. This generalizes traditional scalar sensitivity parameters and accommodates rich parametric and semi-parametric models, enabling more flexible, exposure-specific bounds on the dose-response curve. Moreover, replacing a uniform sensitivity scalar with an exposure-specific function yields narrower partial identification bands for the dose–response curve.
    \item We characterize a class of admissible sensitivity functions for Rosenbaum and Marginal sensitivity models, respectively. All functions in this class satisfy naturally occurring model restrictions but otherwise vary freely. Crucially, as we show for any such function, there exists a full-data distribution law compatible with the bound. 
    \item We derive bounds 
    on the potential outcome mean under each model, using a new characterization of a bound for the conditional average counterfactual regression with continuous treatment, a result which may be of independent interest. 
    Specifically, building upon the work of \cite{yadlowsky2022bounds}, we map exposure density ratio constraints to likelihood ratio bounds for potential outcomes, and devise a common framework that can derive bounds under both the Rosenbaum and Marginal models. The key functional under the Rosenbaum model is shown to be related to a relevant expectile (\cite{newey1987asymmetric, waltrup2015expectile}), and under the marginal model to a corresponding quantile and the conditional value-at-risk function (CVar) (\cite{rockafellar2000optimization, rockafellar2006generalized}; see Section \ref{partial-id} for further details).
    \item Next, we derive doubly robust estimators using pseudo-outcomes and semi-parametric techniques from \cite{kennedy2017non} and \cite{yang2023forster}. While the bounds themselves are not pathwise-differentiable under continuous exposures, we derive relevant influence functions for certain marginalized versions of the bounds, which define pseudo-outcomes used to estimate the targeted bounds nonparametrically via flexible machine learning techniques, with certain small-bias guarantees we characterize. 
    \item We establish $L^2$-consistency and asymptotic normality for the bound estimators using non-parametric ordinary least-squares over a dense basis system. The pseudo-outcome construction leads to second order 
    bias terms, enabling minimax-optimal convergence even with slower rates of nuisance estimation. We also establish consistency for a variance estimator, enabling the construction of valid pointwise confidence interval for the dose-response curve.
    \item Next, we discuss the differences between the Rosenbaum and Marginal models in extensive detail. We illustrate a geometric intuition to both models by identifying a set, such that the Rosenbaum model bounds the diameter of that set, while the Marginal model bounds a relevant radius. This also allows us to extend familiar relationships between both models derived in \cite{zhao2019sensitivity} under binary exposures to their continuous counterparts.
    \item Finally, we show that both the Rosenbaum and Marginal models with any sensitivity function from the function class in (ii) are compatible with the observed data distribution, but only the Rosenbaum model goes a step further by being compatible with arbitrary specifications of the marginal distribution of the unmeasured confounder. This distinction motivates a discussion of \textit{global} versus \textit{targeted} sensitivity analysis, and the contexts in which each model is more appropriate.
\end{enumerate}

The remainder of the paper is organized as follows. Section \ref{setup} we introduce the sensitivity analysis setup, and formally introduce the notion of sensitivity functions and the Rosenbaum and Marginal sensitivity models. Section \ref{partial-id} produces bounds on the dose-response curve using a common regression formulation. Section \ref{inflfn} derives the double robust mapping using semi-parametric techniques, while Section \ref{asymp-prop} establishes the $L^2$-consistency and asymptotic normality in the second-step of counterfactual regression. In Section \ref{comparison}, we dive into a comparative discussion between the two models and identify contexts we believe are appropriate for each. Finally in Section \ref{numer}, we evaluate the performance of our methods through simulations, and apply our method to measuring the effect of second hand lead exposure on blood lead levels in children.

\section{Problem Setup} \label{setup}
\textbf{Glossary.} We denote the CDF associated with any random variable $A$ as $F_A$, a conditional CDF of $A|B=b$ by $F_{A|B}(\cdot|b)$, and the conditional quantile function $Q_{A|B=b}(\tau) = \inf\{x\in \R:F_{A|B}(x|b)\ge \tau\}$ The density of a random variable $A$ is denoted by $f_A$, and the analogous notation holds for joint and conditional densities. We also define $a_+ =\max\{a,0\}$, $a_-=-\min\{a,0\}$, and $a_\pm^2 = (a_\pm)^2$, for any real number $a$. For any vector $a\in \R^d$, let $\|A\| = (\sum_{i=1}^d a_{i}^2)^\frac 12$ denote its vector norm. Also, for any random variable $A$ (either scalar or a vector in $\R^d$) let $\|A\|_{p} = (\E \|A\|^p)^{\frac 1p}$, for $p = \R^{\ge 0}\cup\{\infty\}$. We shall generally represent nuisance functions by $h$ and their estimates by $\hat h$, and for any $p$ as above, define $\|A\|_{p|\hat h} = (\E[\|A\|^p|\hat h])^{\frac 1p}$. Also, for any matrix $B\in \R^d\times \R^d$, let $\|B\|$ denote the operator norm, ie, the largest eigen value of $B$. Finally, we write two sequences $\{a_k\}_1^\infty$ and $\{b_k\}_1^\infty$ to be $a\lesssim b$ if $\exists c>0$ and $K>0$ such that $a_k\le Cb_k$ for $k\ge K$, and $a\simeq b$ if $a\lesssim b$ and $b\lesssim a$; and for random variables $A$ and $B$, we denote $A\lesssim_P B$ if $A = O_p(B)$, and $A\simeq B$ if $A \lesssim_P B$ and $B\lesssim_P A$.


We assume access to $n$ independent and identically distributed samples of $(W_1,\cdots, W_n)$, where $W = (X, T, Y)$, with $X\in \mathcal{X}$ denoting a rich set of measured confounders, $T\in \mathcal{T}$ is a continuous exposure (or treatment, used interchangeably), and $Y\in \mathcal{Y}$ is the outcome of interest. We adopt the potential outcome framework to delineate our causal quantities, denoting $Y(t)$ as the potential outcome that would have been observed had $T$ been set to $t$ (\cite{neyman1923application,rubin1974estimating}) 
Throughout this paper, we are interested in the counterfactual quantity $\{\E[Y(t)]\}_{t\in\mathcal{T}}$, referred to as \emph{the average dose-response curve}.

Suppose $T|X$ admits a density with respect to some base measure $\lambda_b$. We make two standard causal assumptions of consistency and positivity. 
\begin{assumption} \label{std}
        (i) Consistency: $Y(T) = Y$.
        (ii) Positivity: $f_{T|X}(t|x)\ge p_{\m} > 0 \ \ \forall x\in \mathcal{X}$.
\end{assumption}
The consistency assumption bridges the observed outcomes $W$ to the counterfactual quantity of interest $Y(\cdot)$, and implicitly assumes away complex situations like interference and spillover effects across units, and multiple versions of the same treatment. 
While the no-unmeasured confounding (NUC) assumption, alluded to in Section \ref{intro} and formalized as $\{Y(t)\}_{t\in\T}\indep T|X$, in conjunction with Assumption \ref{std} allows us to identify the potential outcome mean, NUC is generally not plausible in observational studies. Instead, one may imagine an unmeasured confounder $U\in\mathcal{U}$ such that, one may generalize the confounding condition, as in Assumption \ref{uc}.
\begin{assumption}[Latent Ignorability] \label{uc}
    $\{Y(t)\}_{t\in\mathcal{T}}\indep T|X, U.$
\end{assumption}
This is a generalization of the standard no-unmeasured confounding assumption (also known as treatment exchangeability or ignorability in the causal inference literature), whereby, while conditioning on $X$ alone does not ensure exchangeability, further conditioning on $U$ does. Assumption \ref{uc} is quite mild, since $U$ is not measured; however, the counterfactual dose response curve is no longer identifiable under the condition, without a different condition. 


To address unmeasured confounding by $U$, one strategy is to impose additional structure on their distribution, possibly involving auxiliary variables have been measured, such as instrumental variables or proxies, which may, under stringent conditions, lead to point identification of the counterfactual dose response curve (\cite{heckman2010comparing, newey2003instrumental, miao2018identifying}). 
In contrast, we adopt a different approach: we aim to partially identify the potential outcome mean under interpretable assumptions about the relationship between the unmeasured confounders and the exposure. A common strategy is to specify a sensitivity model, following the conceptual framework introduced by Rosenbaum (\cite{rosenbaum2002observational}). In this work, we consider a generalized version of such a model that accommodates continuous exposures.

\begin{assumption}[Rosenbaum sensitivity model] \label{rosen-model}
    For a specified $\Gamma:\mathcal{T}\times \mathcal{T}\to [1,\infty)$ satisfying $\Gamma_{t',t} = \Gamma_{t,t'}$ and $\Gamma_{t,t} = 1$, 
    the law of $(T,X,U)$ satisfies 
    \begin{equation} \dfrac{1}{\Gamma_{t,t'}}\le \dfrac{f_{T|U,X}(t|u, X)/f_{T|U,X}(t|\tilde u, X)}{f_{T|U,X}(t'|u, X)/f_{T|U,X}(t'|\tilde u, X)}\le \Gamma_{t,t'} \text{ a.s. }X,\ \ \ \forall t, t'\in\mathcal{T}; u, \tilde u\in\mathcal{U}.
    \label{rosen-eq}
    \end{equation}
\end{assumption}
Assumption \ref{rosen-model} bounds the ratio of conditional densities of $T$ given $X$ and two distinct values of $U$ by a prespecified function of $t$. 
Note that taking $\Gamma_{t,t'} = 1$ is equivalent to no unmeasured confounding, since the conditional density of $T|U=u,X$ would no longer depend on $u$. In case of binary $T\in\{0,1\}$, replacing the conditional density of $T$ by its natural conditional probability analogue yields the classical Rosenbaum sensitivity model on the odds ratio scale, with $\Gamma_{1,0} = \Gamma \ge 1$. In fact, the model imposed by Assumption \ref{rosen-model} can be considered a generalization of the sensitivity model with many-ordered treatments introduced in \cite{rosenbaum1989sensitivity}. It is worth emphasizing that, although we follow the binary-treatment literature in omitting $X$ from the sensitivity function $\Gamma_{t,t'}$, our theory and methods extend directly to sensitivity functions that also depend on $X$, without requiring any additional technical assumptions.

Another relatively more recent sensitivity model is the Marginal sensitivity model of \cite{tan2006distributional}, also see \cite{zhao2019sensitivity, dorn2023sharp, dorn2024doubly}, introduced in the context of binary exposures. A natural generalization  for continuous exposures  entails:

\begin{assumption}[Marginal sensitivity model]\label{mar-model}
    For a specified $\Lambda:\T\times \T\to [1,\infty)$ symmetric in its arguments, with $\Lambda_{t,t} = 1\ \forall t\in \T$, the law of $(T,X,U)$ satisfies
    \begin{equation}
        \dfrac{1}{\Lambda_{t,t'}}\le \dfrac{f_{T|U,X}(t|u,X)/f_{T|X}(t|X)}{f_{T|U,X}(t'|u,X)/f_{T|X}(t'|X)} \le \Lambda_{t,t'} \ \forall t,t'\in \T; u\in \mathcal{U}. \label{mar-eq}
    \end{equation}
\end{assumption} 

Similar to Assumption \ref{rosen-model}, the Marginal Sensitivity Model in Assumption \ref{mar-model} also bounds the ratio of density ratios, but while the Rosenbaum model compares the density ratios between different strata of the unmeasured confounder $U$, the Marginal model on the other hands compares the full-data conditional density ratio to its marginal observed data counterpart
. We provide a detailed discussion of the merits and possible use-cases of each framework in Section \ref{comparison}.

It is important to note that, unlike usual sensitivity models, both Assumption \ref{rosen-model} and Assumption \ref{mar-model} are not parametrized by a scalar sensitivity parameter, but rather a general \textit{sensitivity function}. Replacing a scalar sensitivity parameter with a sensitivity function as ours allows several standard parametric and semi-parametric models to be considered -- for instance, if $U$ has a compact support and $(A,Y)|X,U$ has a multivariate normal distribution, then no scalar parameter exists that satisfies the bounds of either Equation \eqref{rosen-eq} or \eqref{mar-eq}. One could in principle potentially produce scalar bounds by truncating the exposure level, but this approach may not be appropriate for unbounded exposure, and even if appropriate, the approach could yield unnecessarily wide confidence intervals, by replacing a naturally varying sensitivity function with a uniform worst-case bound. However to make our models useful, we need to obtain a rich class of sensitivity functions well-suited to our goal. In Proposition \ref{expfamily} we achieve this goal, and we describe a rich class of admissible sensitivity functions which are provably compatible with Assumptions \ref{rosen-model} and \ref{mar-model}, from which an analyst can draw good candidates for conducting a given sensitivity analysis.
\begin{proposition} \label{expfamily}
    Consider the set of functions 
    $$\F = \left\{\left.\max\left\{\dfrac{\Upsilon_t}{\Upsilon_{t'}},\dfrac{\Upsilon_{t'}}{\Upsilon_t}\right\}\right| \Upsilon:\mathcal{T}\mapsto (0,\infty) \text{ is an arbitrary measurable function.}\right\}$$
    \begin{itemize}
        \item[(i)] For any function $\Gamma_{t,t'}\in\mathcal{F}$, there exists a law $(T,U,X)$ that satisfies the Rosenbaum model in Assumption \ref{rosen-model} with parameter $\Gamma_{t,t'}$.
        \item[(ii)] For any function $\Lambda_{t,t'}\in\mathcal{F}$, there exists a law $(T,U, X)$ that satisfies the Marginal model in Assumption \ref{mar-model} with parameter $\Lambda_{t,t'}$.
    \end{itemize}
\end{proposition}

The proof of Proposition \ref{expfamily}, as well as various potential options of functions $\Upsilon_t$ to generate $\Gamma_{t,t'}$ is discussed in Appendix \ref{setup-proof} and Table \ref{Sens-Function_tab}.

\section{Partial identification of potential outcome mean} \label{partial-id}

Although Assumption \ref{rosen-model} effectively restricts the degree of unmeasured confounding bias, it does not in of itself provide point-identification of the potential outcome mean. The key identification challenge stems from  the fact that:
$$\E[Y(t)] = \E_X\left[\int\E[Y(t)|T=t',X] \,dF_{T'|X}(t'|X)\right],$$
but since $Y(t)$ is not observed unless $t'=t$, the above expression cannot be identified under Assumptions \ref{std} and \ref{uc}. 
\cite{yadlowsky2022bounds} and \cite{dorn2023sharp} considered the Rosenbaum model and the Marginal model respectively for binary treatments, and derived partial identification strategies for the potential outcome mean. In this section, we extend their results to the case of continuous exposures. To simplify the exposition, our theoretical developments primarily focus on identifying lower bounds for $\E[Y(t)]$ under the Rosenbaum and Marginal models, respectively; corresponding upper bounds can be be obtained by analogy, upon redefining $Z(t) := -Y(t)$ and obtaining the negative value of the infimum of $\E[Z(t)]$ over all counterfactual distributions under the respective sensitivity models.

Let $\theta_{t'}(x,t) := \inf\{\E_Q[Y(t)|T=t',X=x]: Q\in \mathcal{R}_x\}$, where $\mathcal{R}_x$ is the set of all laws of $(\{Y(t)\}_{t\in\mathcal{T}}, T, U)$ conditional on $X=x$ satisfying Assumptions \ref{std}, \ref{uc} and the Rosenbaum model \ref{rosen-model}. Theorem \ref{id-rosen} provides a formal characterization of $\theta_{t'}(x,t)$ in terms of the observed data on which our empirical approach shall be based. 

\begin{theorem}[Partial identification under Rosenbaum Sensitivity Model]
\label{id-rosen}
    
    Define $\psi_\theta^{t,t'}(y):= (y-\theta)_+ - \Gamma_{t,t'}(y-\theta)_-$. If $|\theta_{t'}(x,t)|<\infty$, then $\theta_{t'}(x,t)$ satisfies 
    $$\theta_{t'}(x,t) = \sup\{\mu\in \R: \E[\psi^{t,t'}_\mu(Y(t))|T=t,X] \ge 0\}. $$
    Moreover, if $Y|T=t,X$ has a positive density a.s. on the convex hull of its support, then $\theta_{t'}(x,t)$ is the unique solution to  $$\E\cbrace{\psi^{t,t'}_{\theta_{t'}(x,t)}(Y(t))}{T=t,X} = 0.$$
    Furthermore, $\theta_{t'}(x,t)$ is also the unique (up to measure-zero sets) minimizer of the optimization
    \begin{equation} \label{mom-rosen} \arg\min_g \E[(Y- g(X,t))^2_+ + \Gamma_{t,t'}(Y-g(X,t)))^2_-|T=t,X], \end{equation}
    provided $\E[(Y-\theta_{t'}(X,t))^2_+ + \Gamma_{t,t'}(Y-\theta_{t'}(X,t))^2_-|T=t,X] <\infty.$
    
\end{theorem}

\begin{remark}
\normalfont
    For any $t'$ fixed, $\theta_{t'}(x,t)$ solves the following marginal minimization problem as well:
    $$\min_g \E[(Y- g(X,T))^2_+ + \Gamma_{T,t'}(Y-g(X,T)))^2_-].$$
    This is a simple consequence of the interchange of integral and infimum for normal integrands (\cite{rockafellar2009variational}, Theorem 14.60). Hence one may consider the marginal optimization problem, simpler than its conditional counterpart in Theorem \ref{id-rosen}, when estimating $\hat\theta_{t'}(x,t)$ empirically, for continuous exposure $T$. 
\end{remark}
Theorem \ref{id-rosen} produces partial identification of $\inf_{Q\in\mathcal{R}_x}\E[Y(t)]$ by marginalizing over $\theta_{T}(X,t)$, which is identified by the conditional expectile of the distribution $Y|T=t,X$ 
 as a function of $\Gamma_{t,t'}$. Interestingly, expectiles have recently emerged in the literature as an alternative to quantiles (\cite{philipps2022interpreting}), and the result in Theorem \ref{id-rosen}, analogous to the case of binary treatments in \cite{yadlowsky2022bounds}, underscores the importance of expectiles as a critical causal quantity under sensitivity models. The theorem, proved in Appendix \ref{id-rosen-proof}, is the consequence of a key restriction on the likelihood ratio of $L_{t,t'}(y,x) = \frac{d\P(Y(t)|T=t',X)}{d\P(Y(t)|T=t,X)}$, given by $L_{t,t'}(y,x)\le \Gamma_{t,t'}L_{t,t'}(\tilde y,x)$ for all $y,\tilde y$, which is imposed by the Rosenbaum Model \eqref{rosen-eq}. 

Analogously, the Marginal model also imposes a restriction on the likelihood ratio, given by $L_{t,t'}(y,x) \in [\Lambda_{t,t'}^{-1},\Lambda_{t,t'}]$. Let us denote $\zeta_{t'}(x,t) := \inf\{\E_Q[Y(t)|T=t',X=x]: Q\in \mathcal{M}_x\}$, where $\mathcal{M}_x$ is the set of all full data laws conditional on $X=x$ that satisfy Assumptions \ref{std}, \ref{uc} and the Marginal model \ref{mar-model}. Then the aforesaid restriction imposed by the marginal model provides an identification of $\zeta_{t'}(x,t)$, as illustrate in Theorem \ref{id-mar} and proved in Appendix \ref{id-mar-proof}.

\begin{theorem}[Partial Identification under Marginal Sensitivity Model]
\label{id-mar}
    
    
    Define $\rho^{t,t'}_\mu(y) := \Lambda_{t,t'}^{-1}(y-\mu)_+ - \Lambda_{t,t'}(y-\mu)_-.$ If $|\zeta_{t'}(x,t)|<\infty$, then $\zeta_{t'}(x,t)$ is of the form
    \begin{equation} \label{mom-mar} \zeta_{t'}(x,t) = q_{t'}(x,t) + \E[\rho_{q_{t'}(x,t)}^{t,t'}(Y)|T=t,X=x], \end{equation}
    where $q_{t'}(x,t)=Q_{Y|X=x,T=t}((\Lambda_{t,t'}+1)^{-1})$, ie, the $(\Lambda_{t,t'}+1)^{-1}$-th quantile of $Y|X=x,T=t$.
    
\end{theorem}

\begin{remark} \normalfont 
If $Y|T=t,X=x$ is continuous, a simple calculation reveals that 
\begin{equation}
    \zeta_{t'}(x,t) = \Lambda_{t,t'}\E[Y|T=t,X=x] - (\Lambda_{t,t'}-1)\E[Y|Y>q_{t'}(x,t),T=t,X=x]. \label{simpler}
\end{equation}
\end{remark}

Thus if the above condition holds, Equation \eqref{simpler} could be used to estimate $\zeta$ as a linear combination of the conditional mean, and the truncated tail mean function $\E[Y|Y>q_{t'}(X,T),T,X]$ function, sometimes referred to as the Conditional Value-at-Risk (CVar) function (\cite{rockafellar2000optimization}). It is interesting to note that Theorem \ref{id-mar} recovers the adversarial regression formulation of \cite{dorn2024doubly} in the case of continuous exposure, despite our very distinct proof strategies. Note that both Theorem \ref{id-rosen} and \ref{id-mar} use outcome regression formulations to identify the estimand of interest, which is a convenient strategy for continuous exposures. An adversarial propensity score construction strategy, like that in \cite{dorn2023sharp}, is difficult to apply in the continuous exposure setting, due to the appearance of Dirac-delta at $t$ terms when attempting to partially identify $\E[Y(t)]$. Some recent work has considered kernel based localization to circumvent this issue (eg: \cite{colangelo19052025}) in the context of NUC; but we defer exploring such an approach in our setting to future work. Although we emphasize that while the identification strategies in Theorems \ref{id-rosen} and \ref{id-mar} bypass the need to incorporate propensity scores (ie, conditional density in this case), they still play a key-role in efficiently estimating the estimands defining our bounds, as highlighted in the following sections. 

 \section{Debiased Mapping through pseudo-outcome formulation} \label{inflfn}

Let $r(t) = \E[\theta_T(X,t)]$ and $m(t) = \E[\zeta_T(X,t)]$ be the lower bounds on $\E[Y(t)]$ under the Rosenbaum and Marginal models respectively. If $r(t)$ and $m(t)$ were parametrized by finite-dimensional structures, standard tools from semi-parametric efficiency theory (\cite{van2000asymptotic, tsiatis2006semiparametric, hines2022demystifying}) could be used to derive their efficient influence functions (EIFs). However, under milder smoothness conditions, these functionals are no longer pathwise-differentiable if $T$ is absolutely continuous with respect to the Lebesgue measure, and no $\sqrt{n}$-consistent estimators can exist (\cite{bickel1993efficient, diaz2013targeted}). 

To address this challenge, one can adopt a pseudo-outcome approach analogous to \cite{kennedy2017non,yang2023forster, chernozhukov2024conditional}. 
We define $\psi_R = \E[r(T)]$ and $\psi_M = \E[m(T)]$, where the outer expectations are taken with respect to the marginal distribution $F_T$ of the observed exposure. Unlike $r(t)$ and $m(t)$, the population level functionals $\psi_R$ and $\psi_M$ are now both pathwise-differentiable under standard regularity conditions and therefore admit influence functions under an appropriately defined semiparametric model, which can in turn be used to define so-called pseudo-outcomes as corresponding un-centered influence functions. The key insight of pseudo-outcomes is that under conditions of Theorem 2 of \cite{yang2023forster}, the influence function-based estimators of $\psi_R$ and $\psi_M$ inherit a conditional second-order bias property even for their conditional counterparts, such as $r(t)$ and $m(t)$, thereby providing improved estimators for such nonregular, i.e. non $\sqrt{n}$-estimable,  functionals.

\begin{theorem}[EIF under Rosenbaum sensitivity model]
    \label{rosen-eif}
    Suppose the identifying moment equation in Equation \eqref{mom-mar} holds. Then, the efficient influence function for $\psi_R$ in the nonparametric model for the observed data is given by 
    $$\dfrac{f_T(T)}{f_{T|X}(T|X)}\int \dfrac{\psi_{\theta_{t'}(X,T)}^{T,t'}(Y)}{\nu_{t'}(X,T)}dF_{T|X}(t'|X) + A_R(T,X) +B_R(T) - 2\psi_R,$$
    where $A_R(T,X) = \int\theta_T(X,t) \,dF_T(t)$, $B_R(T) = \int\int \theta_{t'}(X,T)\,dF_{T|X}(t'|x)\,dF_X(x)$, and $$\nu_{t'}(X,t) = \P(Y>\theta_{t'}(X,t)|X,T=t) + \Gamma_{t,t'}\P(Y\le \theta_{t'}(X,t)|X,T=t).$$
\end{theorem}

    
\begin{theorem}[EIF under Marginal sensitivity model] \label{mar-eif}
    Suppose the $Q_{Y|T=t,X=x}(\Lambda_{t,t'}+1)^{-1}$ is a continuity point of $Y|T=t,X=x$ for all $t\in\mathcal{T},x\in\mathcal{X}$. Then, the efficient influence function for $\psi_M$ in the nonparametric model for the observed data is given by 
    $$\dfrac{f_T(T)}{f_{T|X}(T|X)}\int \left[\rho^{T,t'}_{q_{t'}(X,T)}(Y) + q_{t'}(X,T) - \zeta_{t'}(X,T)\right]dF_{T|X}(t'|X) + A_M(T,X) + B_M(T) - 2\psi_M,$$ where $q_{t'}(X,T) = Q_{Y|X,T}((\Lambda_{T,t'}+1)^{-1})$ is the conditional quantile as obtained in Theorem \ref{id-mar}, $A_M(T,X) = \int \zeta_T(X,t) dF_T(t)$ and $B_M(T) = \int\int \zeta_{t'}(X,T) dF_{T|X}(t'|x)dF_X(x)$
\end{theorem}


With the EIFs for $\psi_R$ and $\psi_M$ at hand, we construct pseudo-outcomes based on the recipe from \cite{yang2023forster} and \cite{dalal2024anytime}, as the estimated uncentered EIF, which we fit to produce counterfactual regression estimates of $\hat r(t)$ and $\hat m(t)$ by deploying the following two-step procedure:

\textbf{Step 1}: Divide the dataset into three non-overlapping parts $I_1, I_2$ and $I_3$. From $I_3$, we estimate the relevant nuisance parameters $\hat\theta$, $\hat\nu$, $\hat f(t|x)$ for $\hat r(t)$, and $\hat q$, $\hat \alpha$ and $\hat f(t|x)$ for $\hat m(t)$. 

\textbf{Step 2}: For $i\in I_1$, construct the following pseudo-outcomes
\begin{align*}
    \hat Y_r(W_i) &= \dfrac 1{|I_2|}\sum_{j\in I_2} \hat\theta_{T_j}(X_j,T_i) + \dfrac{\dfrac{1}{|I_2|}\sum_{j\in I_2} \hat f(T_i|X_j)}{\hat f(T_i|X_i)}\int \dfrac{\psi^{T_i,t'}_{\hat\theta_{t'}(X_i,T_i)}(Y_i)}{\hat\nu_{t'}(X_i,T_i)}\hat f(t'|X_i)dt', \\
    \hat Y_m(W_i) &= \dfrac 1{|I_2|}\sum_{j\in I_2} \hat\zeta_{T_j}(X_j,T_i) \\ &\hspace{2em}+ \dfrac{\dfrac{1}{|I_2|}\sum_{j\in I_2} \hat f(T_i|X_j)}{\hat f(T_i|X_i)}\int [\rho^{T_i,t'}_{\hat q_{t'}(X_i,T_i)}(Y_i) + \hat q_{t'}(X_i,T_i) - \hat\zeta_{t'}(X_i,T_i)]\hat f(t'|X_i) dt',
\end{align*}
where the integrals are computed numerically.

\textbf{Step 3}: For $(W,T)\in I_1$, regress $\hat Y_r(W) \sim T$ and $\hat Y_m(W)\sim T$ using non-parametric ordinary least squares (OLS) on a user-specified basis system, to obtain $\hat r(t)$ and $\hat m(t)$. 



We elaborate further on the non-parametric OLS in Section \ref{asymp-prop}. 
It is worth mentioning that cross-fitting rather than sample-splitting could be used to improve efficiency. Specifically, one can divide the dataset into $K\ge 3$ non-overlapping parts $J_1,\cdots, J_K$, and setting $I_2 = J_{i_2}$, $I_3 = J_{i_3}$ and $I_1 = \{1,\cdots, n\} - I_2 - I_3$, where $\{i_1,i_2,\cdots, i_K\}$ is a permutation of $\{1,\cdots, K\}$. The ensuing $\binom K2$ estimates could then be averaged to obtain analogous theoretical guarantees on the averaged estimators with full-sample efficiency.




\section{Estimation and Inference} \label{asymp-prop}

In this section, we elaborate on the non-parametric OLS regression of Step 2 in Section \ref{inflfn}, and provide theoretical guarantees for the convergence of $\hat r(t)$ and $\hat m(t)$ to their respective population analogues. 

Let $\lambda$ denote the Lebesgue measure, and let $\Psi = \{\phi_1(\cdot) \equiv 1, \phi_2(\cdot),\phi_3(\cdot),\cdots \}$ be a sequence of functions such that linear combination of these functions are dense in $L_2(\lambda)$. Examples of such sequences include polynomial series, Fourier series, regression splines (\cite{huang2003asymptotics}), local polynomial partition series (\cite{cattaneo2013optimal}), wavelet series, etc. The idea is to use these basis functions to approximate any $L^2$ function from the sample at hand. For any $J\ge 1$, let $\bar \phi_J = (\phi_1,\cdots, \phi_J)$ be a collection of the first $J$ basis functions. Also, for any $f\in L^2(\lambda)$, define $E_J^\Psi(f):= \inf_{b\in \R^J}\|f - b^T \bar \phi_J\|_{L^2(\lambda)}$ as the error of approximating $f$ by the first $J$ functions of $\Psi$. The choice of a dense basis ensures that $\E_J^\Psi(f)\to 0$ as $J\to\infty$ for any $f\in L^2(\lambda)$. 

Let $Q = \E[\bar\phi_J(T)\bar \phi_J(T)^T]$, and let $\hat Q_{|I_1|} = \frac{1}{|I_1|}\sum_{i\in I_1} \bar \phi_J(T_i)\bar\phi_J(T_i)^T$ be its empirical counterpart in $I_1$. The non-parametric OLS estimator for $a(t)$, $a\in \{r,m\}$ is given by
    $\hat a_J(t) := \bar \phi_J(t)^T\hat\beta_a,$
where 
$$\hat\beta_a :=\arg\min_{b\in \R^J} \dfrac{1}{|I_1|}\sum_{i\in I_1} (\hat Y_a(W_i) - b^T\bar \phi_J(T_i))^2 = \hat Q_{|I_1|}\cdot\dfrac{1}{|I_1|}\sum_{i\in I_1} \bar\phi_J(T_i)\hat Y_a(W_i), \ a\in \{r,m\}.$$

To consider the theoretical properties of this estimator, let us define $\xi_J :=\sup_{t\in\mathcal{T}} \| \bar\phi_J(t) \|$, which plays a key role in the choice of $J$. When $\mathcal{T}$ is compact, for polynomial bases, we have $\xi_J \lesssim J$, whereas for bases like the Fourier series, splines, wavelets, local polynomial partitions etc., $\xi_J \lesssim \sqrt{J}$ (\cite{belloni2015some}). We establish our results under the following technical conditions.

\begin{assumption} \label{tecn}
\begin{itemize}
    \item[(i)] $\exists 0<l_{\min}\le l_{\max}<\infty$ such that the eigenvalues of $Q$ are bounded above by $l_{\max}$ and below by $l_{\min}$. 
    \item[(ii)] $\E[\hat Y_a(W)^2|T]\le \bar\sigma^2_a<\infty $ a.s. $T$, for $a\in \{r,m\}$.
    \item[(iii)] $\xi_J^2\log J/|I_1| \to 0$. 
    \item[(iv)] $\mathcal{T}\subseteq \R^d$ is compact and $T$ has a density $f_T$ on $\mathcal{T}$ with respect to $\lambda$, and $f_T(t)$ is bounded above for $t\in\T$. 
    \item[(v)] $Y|T=t,X=x$ admits a bounded density with repect to the Lebesgue measure. 
\end{itemize}
    
\end{assumption}

Assumption \ref{tecn} (i) ensures that the components of $\bar \phi_J$ are not too co-linear, while maintaining that any linear combination of $\bar \phi_J$ maintains a bounded variance. Such a stability condition can be simply ensured if the components of $\Psi$ are orthonormal on $(\mathcal{T},\lambda)$, and $dF_T/d\lambda$ is bounded above and away from zero (\cite{belloni2015some}). (ii) is a mild regularity condition required to quantify the $L^2$-error of estimation. (iii) quantifies the rate of growth of $J$ in comparison to the sample size. Condition (i) and (iii) allows $\hat Q_{|I_1|}$ to concentrate around $Q$, and hence obtain necessary rates of consistency. We note that the requirement that $\xi_J^2\log J/|I_1|$ could be relaxed to a lesser degree using potentially different learners than the OLS, for instance the Forster-Warmuth learner considered in \cite{yang2023forster}. Here, we primarily focus on the OLS regression due to its familiarity and well-established theoretical properties. 
Condition (iv) allows us to express the $L^2$ rate of convergences in terms of the approximation bias $E_J^\Psi$ with respect to the Lebesgue measure. 

Condition (v) guarantees sufficient smoothness of the mapping $\hat\theta_{t'}(X,T)\mapsto \E[\psi^{T,t'}_{\theta_{t'}(X,T)}(Y)|X,T]$ or that of $q_{t'}(X,T)\mapsto \E[\rho^{T,t'}_{q_{t'}(X,T)}(Y)|X,T]$. On closer inspection, the proof of forthcoming Theorems \ref{l2-cons} and \ref{asymp-norm} show that Condition \ref{tecn} (v) could be relaxed: if $\theta_{t'}(X,T)$ and $q_{t'}(X,T)$, alongwith their estimates, have ranges $\mathcal{A}_r$ and $\mathcal{A}_m$ respectively, then one could instead have ${\text{ess}\sup}_X \sup_{y\in \mathcal{A}_a(X)} f_{Y|T,X}(y|t,x) <\infty,$ which is satisfied when $Y$ is binary and $0<\P(Y = 1|T,X)<1$ for $y\in\{0,1\}$, as the estimates $\hat\theta$ and $\hat q$ will eventually be inside $(0,1)$ and $Y$ would have no mass on any Borel set not containing 0 or 1.

\begin{theorem}[$L^2$-rate of convergence] \label{l2-cons}
    Suppose Assumption \ref{std} and \ref{tecn} hold.  Then, 
    \begin{align}
        \|\hat r_J-r\|_{2|\hat h} &\lesssim_P \sqrt{\dfrac{\bar\sigma^2_r J}{|I_1|}} + E_J^{\Psi}(r) + \|\hat\theta_T(X,T') - \theta_T(X,T')\|^2_{4|\hat h} \nonumber \\ &\hspace{1em} + \|\hat\theta_T(X,T') - \theta_T(X,T')\|_{4|\hat h}\left(\|\hat f(T|X) - f(T|X)\|_{4|\hat h} +  \|\hat\nu_{T}(X,T') -\nu_{T}(X,T')\|_{4|\hat h}\right) \label{rosen-l2}\\
        \|\hat m_J - m\|_{2|\hat h} &\lesssim_P \sqrt{\dfrac{\bar\sigma^2_m J}{|I_1|}} + \E_J^\Psi(m) + \|\hat q_T(X,T') - q_T(X,T')\|^2_{4|\hat h} \nonumber  \\ &\hspace{1em} + \|\hat\zeta_T(X,T') - \zeta_T(X,T')\|_{4|\hat h}\|\hat f(T|X) - f(T|X)\|_{4|\hat h}. \label{mar-l2}
    \end{align}
    where $(T,X)\sim F_{T,X}$ and $T'\sim F_T$, with $T'\indep (T,X)$ on the right hand side of Equations \eqref{rosen-l2} and \eqref{mar-l2}.
\end{theorem}
\begin{remark} \label{rem3}
    Note that in Equations \eqref{rosen-l2} and \eqref{mar-l2}, $J$ appears only in the first two terms. Thus, for $a\in \{r,m\}$, if $\{\eta_j\}_{j=1}^\infty$ constitutes a non-increasing sequence such that $a\in \{f: L_2(\lambda): \E_j^\psi(f) \le \eta_j \ \forall j\ge 1\}$, then, for $J_{|I_1|} = \min\{J: \eta^2_J\le \bar\sigma^2_a J/|I_1|\}$, one can replace the first two terms of Equations \eqref{rosen-l2} and \eqref{mar-l2} and replace them by $\sqrt{J_{|I_1|}/|I_1|}.$
\end{remark}

The proof of Theorem \ref{l2-cons} is given in Appendix \ref{l2-cons-proof}. Interestingly, one can consider Theorem \ref{l2-cons}, in conjunction with Remark \ref{rem3} in the case of smooth function classes like $s$-H\"older or $s$-Sobolev spaces (refer to Appendix \ref{fnclass} for definitions). Suppose $\Psi$ is such that $\xi_J \lesssim \sqrt{J}$ (Fourier, spline, wavelet, local polynomial series, etc.). For $a\in \{r,m\}$, if $a$ belongs to such an $s$-smooth function class, and $\eta_J = O(J^{-2s/d})$, where $d$ is the dimension of $T$, then choosing $J_{|I_1|} \simeq |I_1|^{\frac{d}{2s+d}}$, and assuming the bias due to the estimation of nuisance functions is $\lesssim_P |I_1|^{-\frac{s}{2s+d}}$ yields, $\|\hat a_J-a\|_{2|\hat h} \lesssim_P |I_1|^{-\frac{s}{2s+1}}$, which matches the oracle minimax rate for this problem. The decay condition on $\eta_J$ holds true for H\"older smooth function classes $\Sigma_d^H(s,L)$ and Sobolev spaces of order $s$ for the bases mentioned above. On the other hand, if one uses bases for which $\xi_J$ grows faster than $\sqrt{J}$, for instance, classical polynomial basis where $\xi_J \lesssim J$, then to allow $J_{|I_1|}$ to simultaneously be of the order of $|I_1|^{d/(2s+d)}$ to achieve the minimax rate, while maintaining $\xi_J^2\log\xi_J/|I_1|\to 0$ requires $d/(2s+d)<1/2$ or $s>d/2$. Thus a stricter tradeoff is required when using polynomial bases with non-parametric OLS.

Theorem \ref{l2-cons} also highlights the utility of using pseudo-outcomes constructed in Section \ref{inflfn}. For any basis dense in $L^2(\lambda)$, taking $J\to\infty$ and $J/|I_1|\to 0$ allows achieving consistency, as long as the nuisance estimation errors vanish suitably. In particular, for $\hat r_J$, it suffices that either $\|\hat \theta_T(X,T') - \theta_T(X,T')\|_{4|\hat h}\to 0$, or $\|\hat f(T|X) - f(T|X)\|_{4|\hat h}\to 0$ and $\|\hat \nu_T(X,T') - \nu_T(X,T')\|_{4|\hat h}$. Similarly, for $\hat m_J$, it is sufficient for either $\|\hat q_T(X,T')-q_T(X,T')\|_{4|\hat h}\to 0$, or $\|\hat\zeta_T(X,T') - \zeta_T(X,T')\|\to 0$ and $\|\hat f(T|X) - f(T|X)\|_{4|\hat h} \to 0$. 
These conditions reflect the lower order bias (more precisely second order bias) built into the pseudo-outcome. In particular, the second order terms appearing in Equations \eqref{rosen-l2} and \eqref{mar-l2} help facilitate the attainment of minimax rates: they allow the estimation errors due to nuisance functions to converge at the rate $ \|I_1\|^{-\frac{s}{2s+d}}$, even when the nuisance functions themselves are less smooth than the target curves $r(t)$ and $m(t)$.

Next, we focus on pointwise limit theory for $\hat r(t)$ and $\hat m(t)$, and establish pointwise asymptotic normality results. For that, define $Y_a(W)$ as the oracle analogue of $\hat Y_a(W)$ in Step 2 of Section \ref{inflfn}; in which, we replace nuisance function estimators $\hat h$ by their true counterparts, and let $\varepsilon_a = Y_a(W) - a(T)$. Then, we formulate Assumption \ref{tecn-2} required to achieve pointwise asymptotic normality. 
\begin{assumption} \label{tecn-2}
For $a\in \{r,m\}$, let the following hold.
    \begin{itemize}
        \item[(i)] $\sup_{t\in \T} |a(t)|<\infty$. 
        \item[(ii)] $\exists \delta>0$ such that $\sup_{t\in \T}\E[|Y|^{2+\delta}|T=t]<\infty$. Furthermore, 
        
        (a) $\sup_{t,t'\in \mathcal{T}}\E[|\theta_{t'}(X,t)|^{2+\delta}|T=t]<\infty$. 
        
        (b) $\sup_{t,t'}\E[|\zeta_{t'}(X,t)|^{2+\delta}|T=t]<\infty$ and $\sup_{t,t'\in\T}\E[|q_{t'}(X,t)|^{2+\delta}|T=t]<\infty$.
        \item[(iii)] (a) $\V(\theta_T(X,t))>0$ (b) $\V(\zeta_T(X,t))>0$, for all $t\in \T$.
        \item[(iv)] Let $\beta^a_J = \arg\inf_{b\in \R^J} \|Y_a(W) - b^T\bar\phi_J(T)\|_2$, and let $l_J = \sup_{t\in\T} |a(t) - \bar\phi_J(t)^T\beta^a_J|$. Then, $\sqrt{J\xi_J^2\log J/|I_1|} l_J \to 0$. 
    \end{itemize}
\end{assumption}
Assumption \ref{tecn-2}(i) and (ii) are regularity conditions to control the behavior of the target functional $a(t)$ as well as the moments of the nuisance functions. It imposes restrictions on the tails of the regression errors, so that a Lyapunov-type condition can be established to produce normality, even when $J$ varies with sample size. (iii) ensures that the target functionals are non-trivial. (iv) ensures that the pointwise linearization of $\hat a_J(t) - a(t)$ holds under non-parametric OLS, with unknown design matrix $Q$. 
These conditions lead us to the asymptotic normality result we establish in Theorem \ref{asymp-norm}.
\begin{theorem}[Asymptotic normality] \label{asymp-norm}
Suppose Assumptions \ref{std}, \ref{tecn} and \ref{tecn-2} hold. Let $\Omega = Q^{-1}\E[\varepsilon_a^2]Q^{-1}$, and $s(t) = \Omega^{\frac 12}\bar\phi_J(t)$. Suppose $a(t) - \bar\phi_J(t)^T\beta_J^a = o(\|s(t)\|) $.  Furthermore, let the nuisance functions be estimated such that $\xi_J\|\hat h - h\|_{2|\hat h} = o_p(1)$, and 
\begin{itemize}
    \item[(a)]  $\|\hat\theta_T(X,T') - \theta_T(X,T')\|^2_{4|\hat h} = o_p\left(\frac{1}{\sqrt{J|I_1|}}\right)$; $\|\hat\theta_T(X,T') - \theta_T(X,T')\|_{4|\hat h}\|\hat f(T|X) - f(T|X)\|_{4|\hat h} = o_p\left(\frac{1}{\sqrt{J|I_1|}}\right)$ and $\|\hat\theta_T(X,T') - \theta_T(X,T')\|_{4|\hat h}\|\nu_T(X,T') - \nu_T(X,T')\|_{4|\hat h} = o_p\left(\frac{1}{\sqrt{J|I_1|}}\right)$
    \item[(b)] $\|\hat q_T(X,T') - q_T(X,T')\|^2_{4|\hat h} = o_p\left(\frac{1}{\sqrt{J|I_1|}}\right)$ and $\|\hat\zeta_T(X,T') - \zeta_T(X,T')\|_{4|\hat h}\|\hat f(T|X) - f(T|X)\|_{4|\hat h} = o_p\left(\frac{1}{\sqrt{J|I_1|}}\right)$,
\end{itemize}
where $(T,X), T'$ are as in Theorem \ref{l2-cons}. Then, 
$$\dfrac{\sqrt{|I_1|}(\hat a_J(t) - a(t))}{\|s(t)\|}\to\mathcal{N}(0,1).$$
\end{theorem}
The proof of Theorem \ref{asymp-norm} can be obtained in Appendix \ref{asymp-norm-proof}. Note that the rate conditions imposed upon the nuisance function estimates are of second order, as was obtained in Theorem \ref{l2-cons}, again a consequence of the influence function-based pseudo-outcome construction. The condition $a(t) - \bar\phi_J(t)\beta_J^a = o(\|s(t)\|)$ can be perceived of as an undersmoothing condition. Typically, one would expect $s(t) \simeq \sqrt J$, and in that case the undersmoothing condition can be replaced by $l_J = o(\sqrt{J/|I_1|}).$ If one considers $a$ belonging to a H\"older or Sobolev classes of smoothness $s$, then $l_J \lesssim_P J^{-s/d}\log J$, and thus we require $\sqrt{I_1}J^{-\frac{s}{d} -\frac 12}\log J\to 0$.
\begin{theorem}[Estimation of variance] \label{var-est}
Suppose in addition to Assumptions \ref{std}, \ref{tecn}, and \ref{tecn-2}, and the regularity conditions mentioned in Appendix \ref{var-est-proof} hold. Let  $\hat\Sigma = \frac{1}{|I_1|}\sum_{i\in I_1} (\hat Y_a(W_i) - \hat a(T_i))^2$, and $\hat \Omega = \hat Q_{|I_1|}^{-1}\hat\Sigma\hat Q_{|I_1|}^{-1}$. Then, 
$$\|\hat \Omega - \Omega\| \lesssim_P (|I_1|^{\frac{1}{2+\delta}}+l_J)\left(\sqrt{\dfrac{\xi_J^2\log J}{|I_1|}}+\lambda_h\right) + \|\hat h - h\|^2_{\infty|\hat h},$$
where $\lambda_h = \max_{i\in I_1} |\hat h(W_i) - h(W_i)|$.
In particular, the right-hand side is $o_p(1)$ if $\|\hat h- h\|_{\infty|\hat h}\to 0$ and $\lambda_h(|I_1|^\frac{1}{2+\delta} +l_J) = o_p(1)$.
\end{theorem}
Theorem \ref{var-est}, proved in Appendix \ref{var-est-proof}, allows for consistent estimation of the asymptotic variance, and thus $\Omega$ in Theorem \ref{asymp-norm} can be consistently replaced by $\hat\Omega$ to produce pointwise confidence intervals for $r(t)$ and $m(t)$. Uniform confidence bands may also be obtained using a similar procedure under slightly stronger conditions, analogous to Section 4.3 of \cite{belloni2015some}.

A comment is warranted about the estimation of nuisance functions. Despite the dependence on both $t$  and $t'$, the estimation of the expectile $\theta_{t'}(x,t)$ and $q_{t'}(x,t)$ can be considered an $M$-estimation problem, as highlighted in Equation \eqref{mom-rosen} for the former, and an analogous pinball loss-formulation for the latter conditional quantile. Conditions and rates for convergence for such general $M$-estimation problems using sieve estimators have been extensively considered in \cite{chen1998sieve, chen1999improved, chen2007large} (Chapter 76). Learning expectile regression as in $\theta_{t'}(x,t)$ have also been considered using kernel methods (\cite{farooq2019learning}) and support vector machines (\cite{farooq2017svm}). Having obtained $\theta_{t'}(x,t)$ and $q_{t'}(x,t)$, the additional nuisance functions $\nu_{t'}(x,t)$ and CVar could again be formulated as $M$-estimation problems, and sieve-based estimators could in principle likewise be used. 
Estimation for the CVar function has also previously received some attention, for instance see \cite{cai2001weighted}. Estimation of conditional density functions has also been well studied in the literature, for instance one could deploy Nadaraya-Watson estimators (see \cite{wand1994kernel}). In practice, one could also use machine learning to estimate nuisance functions for additional flexibility and potentially improved empirical performance
To implement such methods, we propose dividing $I_3$ into a partition $\{I_3^1, I_3^2\},$ where we learn the first stage nuisance functions $\theta_{t'}(x,t)$ and $q_{t'}(x,t)$ using ML techniques, and learn $\nu_{t'}(x,t)$ and $\zeta_{t'}(x,t)$ from $I_3^2$ and estimates $\hat\theta_{t'}(x,t)$ and $\hat\zeta_{t'}(x,t)$. The integrals required to compute the pseudo-outcomes in $\hat Y_r$ and $\hat Y_m$ may be evaluated numerically using Gaussian quadratures (\cite{stoer1980introduction}).
\section{Competing sensitivity models: which to use?} 
\label{comparison}
%

Both the Rosenbaum and marginal sensitivity models impose restrictions on the ratio of density ratios, which -- unlike the approach in \cite{bonvini2022sensitivity} -- can be interpreted analogously to odds ratios in the binary treatment setting (\cite{chen2007semiparametric, tchetgen2010doubly}).
In contrast, models that \textit{directly} bound the likelihood or density ratios by a scalar—such as those in \cite{bonvini2022sensitivity} and \cite{jesson2022scalable}—impose overly strong restrictions on the conditional distribution of the exposure, often making the assumption difficult to satisfy in practice. Even when the scalar is replaced by a sensitivity function, as in Assumptions \ref{rosen-model} and \ref{mar-model}, the interpretation remains opaque. For example, if the exposure space $\T$ is truncated— as is common for technical reasons in nonparametric regression—the meaning of the sensitivity parameter becomes entangled with the choice of truncation threshold, complicating interpretation. We elaborate on this issue further in Appendix \ref{bkvw}.



In comparing the Rosenbaum and Marginal models, the key distinction is that the former restricts the density ratios for pairs of $u$ and $u'$, whereas the latter is instead constraining the conditional and unconditional distribution of the exposure $T$ within a strata of $X$. There is an interesting geometric interpretation to these models, as delineated in Proposition \ref{geometry}.

\begin{proposition}
    \label{geometry}
    Consider the set $S_{t,t'|X} = \left\{\dfrac{f_{T|U,X}(t|u,X)}{f_{T|U,X}(t'|u,X)}: u\in\mathcal{U}\right\} \in (0,\infty)$, that is the range of the function $g_{t,t'|X}:\mathcal{U} \mapsto \R$ such that $g_{t,t'|X}(u) = f_{T|U,X}(t|u,X)/f_{T|U,X}(t'|u,X)$. Then, 
    \begin{itemize}
        \item[(i)] Equation \eqref{rosen-eq} bounds the diameter of $S_{t,t'|X}$ under (the metric induced by) the norm $|\log(\cdot)|$ by $\log\Gamma_{t,t'}$, where diameter of a set $A$ with metric $d$ is given by $\sup_{x,y\in A} d(x,y)$.
        \item[(ii)] If $\mathcal{U}$ is connected and $g_{t,t'|X}$ is continuous in $u$, then Equation \eqref{mar-eq} implies that there exists a point $u^*$ such that $S_{t,t'|X} \subseteq B(g_{t,t'|X}(u^*), \log \Lambda_{t,t'})$, where $B(a,r)$ is the ball of radius $r$ (with respect to the same metric as before) from the point $a$. 
    \end{itemize}
\end{proposition}

\begin{remark} \normalfont
(i) $\mathcal{U}$ being connected is a relatively mild requirement, since even for discrete latent confounders, one can imagine $\mathcal{U}$ as a refinement of the unmeasured confounder space. (ii) Even if $\mathcal{U}$ is not connected, say in the case of a discrete unmeasured confounder, $f_{T|X}(t|X)/f_{T|X}(t'|X)$ lies in the convex hull of $\mathcal{U}$, and the Marginal model can be thought of as bounding the radius of the set $S_{t,t'}$ from the point $f_{T|X}(t|X)/f_{T|X}(t'|X)$. However, the interpretation to Proposition \ref{geometry} is cleaner under the stated assumptions. 
\end{remark}

Proposition \ref{geometry}, proved in Appendix \ref{geom-proof}, highlights the key distinction between the Rosenbaum and Marginal sensitivity models (analogous distinctions still hold in the case of binary exposures). The Rosenbaum model restricts the diameter of the set characterizing the relationship between the exposure and the unmeasured confounder $U$, whereas the Marginal model instead restricts its `radius' from a reference point $u^*$ (depending on the marginal $T|X$). This also allows us to rediscover familiar relationship between the Rosenbaum and marginal models, as pointed out in \cite{zhao2019sensitivity} (reformulated to accommodate continuous exposures in \ref{corol1}), using simple geometric arguments (Appendix \ref{corol1-proof}). 
\begin{corollary}
\label{corol1}
If the joint distribution of $(T,U,X)$ is such that it satisfies 
\begin{itemize}
\vspace{-1em}
    \item[(i)] the Rosenbaum model in Equation \eqref{rosen-eq} with parameter $\Gamma_{t,t'}$, then it also satisfies the Marginal model in Equation \eqref{mar-eq} with parameter $\Gamma_{t,t'}$.
    \item[(ii)] the Marginal model in Equation \eqref{mar-eq} with parameter $\sqrt{\Gamma_{t,t'}}$, then it also satisfies the Rosenbaum model in Equation \eqref{rosen-eq} with parameter $\Gamma_{t,t'}$.
\end{itemize}
\end{corollary}
The Rosenbaum sensitivity model has another important feature worth emphasizing: it is invariant to the marginal distribution of $U$, since it only specifies a relationship in the treatment mechanism $T|U,X$, and is always compatible with the observed data distribution $(Y,T,X)$, as highlighted in Theorem \ref{data-compatibility}.
\begin{theorem} \label{data-compatibility}
    \begin{enumerate}
        \item For any $\Gamma_{t,t'} 
        \in \F$ (as in Proposition \ref{expfamily}) measurable with respect to $F_{T|X}\times F_{T|X}$,
        and any arbitrary distribution $G$ absolutely continuous wrt a base measure $\mu_{U|X}$ on $\mathcal{U}$, there exists a joint distribution $\Q$ for $(T,X, U, \{Y(t)\}_{t\in \mathcal{T}})$ such that (i) the distribution of the observables $(T,X,Y)$ is the same as under $\P$ and $\Q$, (ii) the distribution of $U|X$ is the same under $\Q$ and $G$, and (iii) $\Q$ satisfies Assumption \ref{rosen-model}. 
        \item For any $\Lambda_{t,t'}\in \F$ measurable with respect to $F_{T|X}\times F_{T|X}$, there exists a joint distribution $\Q$ for $(T,X,U,\{Y(t)\}_{t\in\mathcal{T}})$ such that (i) the distribution of the observables $(T,X,Y)$ is the same as under $\P$ and $\Q$, and (ii) $\Q$ satisfies Assumption \ref{mar-model}. 
    \end{enumerate}
\end{theorem}

\begin{remark}
\label{rem:mar-nonglobal} \normalfont
Part (1) in Theorem \ref{data-compatibility}, could be strengthened to obtain $$\sup_{u,u'\in\mathcal{U}} \dfrac{f_{T|U,X}(t|u,X)f_{T|U,X}(t'|u',X)}{f_{T|U,X}(t'|u,X)f_{T|U',X}(t|u',X)} = \Gamma_{t,t'}\in \F \text{ a.s.},$$  even after allowing for the distribution of $T$ and $U$ to be arbitrary. However, for the marginal model, such a strengthening keeping the distribution of $U$ to be arbitrary, would not be possible - specifying a distribution of $T$ and $U$ necessarily imposes further restrictions on achievable values of $\Lambda_{t,t'}$, as illustrated in Appendix \ref{mar-nonglobal-proof}.
\end{remark}
Theorem \ref{data-compatibility} is related to a result in \cite{osius2009asymptotic}, though we provide a fully self-contained proof in  Appendix \ref{data-compatibility-proof}. The theorem underscores a key distinction of the two models - not only is the Rosenbaum sensitivity model data-compatible, a property it shares with its Marginal counterpart, but it also leaves the marginal distribution of $U$ unrestricted, making it particularly suited for addressing certain sensitivity questions. 

To elaborate upon that point, one can think of a sensitivity analysis to NUC in the following distinct ways - (a) as a global statement\footnote{\cite{leamer1985sensitivity} uses the term ``global sensitivity analysis" in a related but distinct concept; our interpretation in inspired by, but not identical to, his.}: `How robust are the study's conclusions to \textit{any} potential unmeasured confounder?', or (b) as a targeted statement: `How would the results of the study vary if a \textit{specific, plausible} unmeasured confounder would have been included?'. Because the Rosenbaum model is agnostic to the marginal law of $U$, it is well-suited for global statements of type (a), by remaining agnostic to the nature of the confounder and seek a general sense of robustness to causal conclusions. On the other hand, the marginal sensitivity model, as elucidated in Proposition \ref{geometry}, measures the distances from a particular confounder level $u^*$, which depends on the marginal distribution of $U$ (through that of $T$). Thus it may be more informative when one has external information about the unmeasured confounders, either through knowledge of say the prevalence of a binary $U$ in the population, negative control proxies, instrumental variables, etc. In such cases, the marginal model may yield tighter bounds than the Rosenbaum model (since if one believes the Marginal model with $\sqrt{\Lambda}$, it still needs the Rosenbaum model with parameter $\Lambda$), subject to the validity of the implied marginal distribution of $U$. 

The distribution of $U$ could also be informed by other important aspects of a study, like a negative control outcome or an instrumental variable, and hence can thus be incorporated into a marginal model to produce tighter bounds on the estimands of interest. While the Rosenbaum model can still be used in such analyses, the Marginal model with function $\Lambda$ being a subset of the Rosenbaum model is expected to yield tighter bounds, subject to the credibility of the marginal distribution of $U$. 
\section{Numerical illustration} \label{numer}

We next demonstrate the performance of Marginal and Rosenbaum sensitivity models in simulated instances and a case study.

\subsection{Simulation Experiments}
We evaluate the finite-sample performance of both sensitivity models using the following data-generating process:
\begin{align*}
    X &\sim \text{Uniform}[-0.5,0.5]^4\\
    T|X &\sim \text{Beta}(\lambda(X), 1-\lambda(X)), \text{ where }\text{logit}(\lambda(X)) = 0.2X_1 + 0.5X_2 + 0.7X_3\\
    Y|T,X &\sim \mathcal{N}(0.4X_1 + 0.2X_2 + 0.9X_3 + \sin(\pi T) + \sin(2\pi T), (1+X_4^2)(1+T)).
\end{align*}
We are therefore evaluating the methods under NUC conditions.  The distribution for $T$ is similar to that in \cite{kennedy2017non}, and the conditional variance of $Y$ varies with both $X_4$ and $T$ incorporate heteroskedasticity. We used sensitivity functions $\Lambda_{t,t'} = \exp(\log(5)|t-t'|)$ and $\Gamma_{t,t'} = \exp(\log(25)|t-t'|)$, in line with the relationship between the two models from Corollary \ref{corol1}. 

Conditional density is estimated via binning with neural networks, while outcome regression is estimated via \texttt{xgboost}. Quantiles in the Marginal model are fit using natural splines in $X$ and polynomial sieve in $T$. CVar function has been estimated by \text{xgboost}, and $\hat\zeta_{t'}(x,t)$ is constructed via Equation \eqref{simpler}. For the Rosenbaum model, 
expectiles are learned across a grid of $\tau$ values, and the closest $\tau$ to $(1+\Gamma_{t,t'})^{-1}$ has been used to approximate $\theta_{t'}(x,t)$. Both expectiles and conditional tail CDF are estimated by $\texttt{xgboost}$, and $\nu_{t'}(x,t)$ is estimated via plugins of the same. 

We use a sample size of $n= 5000$, split into three parts: roughly 2500 for nuisance estimation, and the rest split in half for the aggregation ($I_2$) and evaluation ($I_1$) sets. The final regression step uses a degree-5 polynomial regression, and the estimates have been cross-fitted to gain efficiency.
\begin{figure}[!ht]
    \centering
    \includegraphics[width=\textwidth]{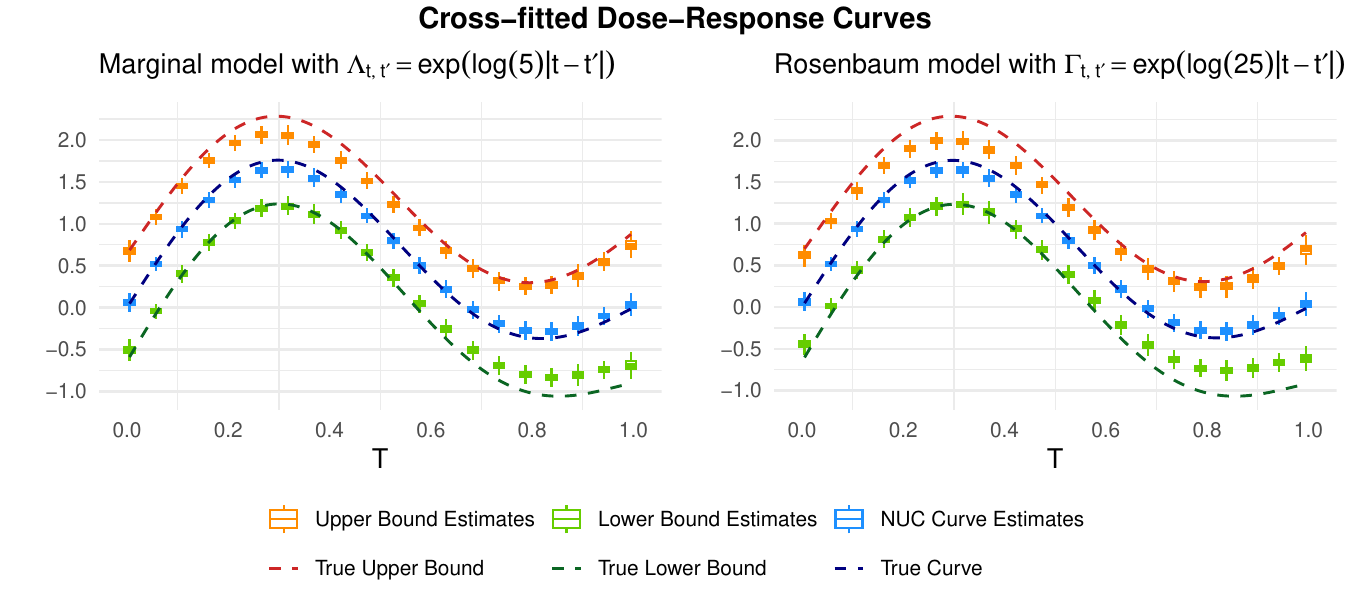}
    \caption{Cross-fitted dose-response curve estimates and sensitivity bounds for Marginal and Rosenbaum Sensitivity models}
    \label{combo}
\end{figure}

The experiment is repeated 500 times, and Figure \ref{combo} shows boxplots of the resulting dose-response and sensitivity bound estimates across a grid of $T$. The NUC estimates (from \cite{semenova2021debiased}), as well as the true-curve are overlayed for reference. 
As expected, the estimated curves generally agree with the truth. However, despite the moderately large sample size, we observe finite-sample smoothing bias in the estimates, which is more pronounced in the bounding curves than in the curve based on NUC. These biases arise from nuisance estimation as well as from the undersmoothing inherent in approximating the curves with polynomial bases.
Overall, the simulations highlight the tradeoff between robustness and flexibility in nonparametric sensitivity analysis. While correct parametric or semiparametric specifications could potentially yield faster convergence rates and thus reduce finite sample bias, they could be inconsistent if incorrect. On the other hand, nonparametric estimation of high-dimensional nuisance functions using machine learning methods generally requires larger sample sizes for reliable convergence -- especially when estimating tail behavior.

\subsection{Application: Effect of second hand smoke exposure on blood lead levels in children}
In this subsection we apply our proposed methods to estimate the effect of second-hand smoke exposure on blood-lead levels in children. While the link between active smoking and elevated blood lead levels is well-established, we focus here on the impact of passive exposure—a critical public health question given the heightened vulnerability of children’s developing nervous systems. Elevated blood lead in children has been linked to reduced intelligence, stunted growth, and anaemia (\cite{national1993measuring}).

\cite{mannino2003second} found an association between second-hand smoke exposure and higher blood lead levels using data from National Health and Nutrition Examination Survey (NHANES). However, the observational nature of the data raises potential concerns about unmeasured confounding. \cite{zhang2020calibrated} addressed this by conducting a sensitivity analysis, finding strong evidence for a causal effect even after accounting for moderate unmeasured confounding. Their analysis used cotinine— a metabolite of nicotine— as a biomarker, dichotomizing its levels to define exposure.

We extend this line of work using more recent NHANES data from 2003–2004 and 2011–2016. Following \cite{mannino2003second}, we restrict to children aged 4-16, and having cotinine levels at most 15.0 ng/mL to exclude active smokers, resulting in 6616 children to be included in our dataset. Measured confounders include poverty-income ratio, age, sex and number of rooms at home. Our method estimates the effect of relative changes in cotinine on blood lead levels. Plausible unmeasured confounders in our analysis includes school type (more active smokers or older lead-based paint in public school buildings \cite{american2005lead}), or source of drinking water (\cite{gibson2020children}). 

\begin{figure}[!ht]
    \centering
    \includegraphics[width=\textwidth]{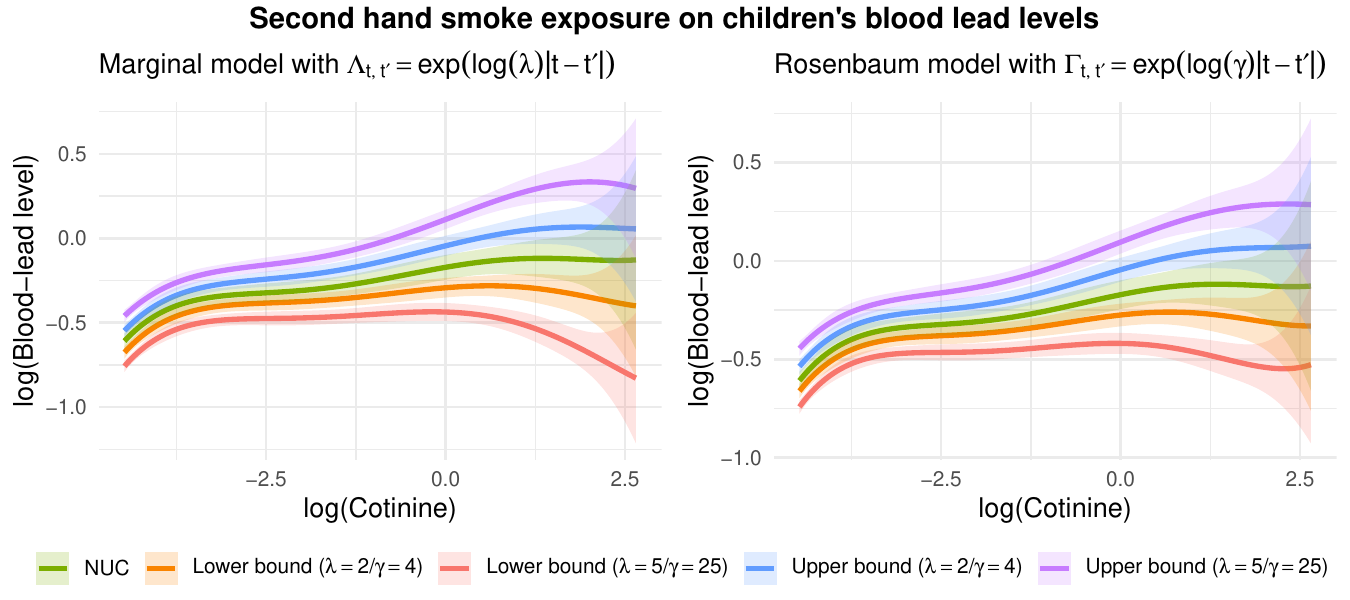}
    \caption{Cross-fitted dose-response curves for percentage change of blood-lead levels for percentage change in cotinine}
    \label{combo_lead}
\end{figure}

Figure \ref{combo_lead} shows a steady rise in log blood lead levels with rising cotinine, consistent with prior findings. For lower cotinine levels, the NUC curve is tightly enveloped by the upper and lower bounds from both the Marginal and Rosenbaum sensitivity models, suggesting robustness to even strong unmeasured confounding. At higher cotinine levels, the bounds begin to widen, possibly due to increased heterogeneity in blood lead levels or smaller sample sizes (reflected in wider standard error estimates) in that exposure range. 

\section{Conclusion}


This work extends the scope of sensitivity analysis in causal inference by introducing generalizations of the Rosenbaum and Marginal sensitivity models to the setting of continuous exposures. Our framework replaces classical scalar sensitivity parameters with sensitivity functions that vary with the exposure level, enabling richer modeling of exposure-confounder relationships and sharper identification bounds for the dose–response curve. In addition, we offer a flexible collection of sensitivity functions—each compatible with the observed data—allowing researchers to tailor their choice based on the contextual demands of the application.

We obtained novel partial identification results for the dose-response curve $\{\E[Y(t)]\}_{t\in\T}$. Our results indicate that both the sensitivity bounds depend on tail behavior of the outcome distribution-- expectiles for the Rosenbaum model and quantiles for the Marginal models. We also establish influence function-based estimators for the sensitivity bounds using a debiased pseudo-outcome formulation, and the resulting estimators are shown to be consistent, achieve the minimax rate, and asymptotically normal under weaker second-order convergence rates of the nuisance functions. 


A common theme we observe across both sensitivity models is that the resulting bounds depend on the tail behavior of the conditional outcome distribution—through expectiles, tail CDFs, quantiles, or CVaR. This matches our intuition --- unmeasured confounding can cause the likelihoods of $Y(t)|T,X$ and $Y|T,X$ to differ markedly, with the worst-case departure quantified by the sensitivity model. The corresponding worst-case likelihood ratio therefore relates to the behavior of the outcome at the extremes, making the functionals tail-dependent.
This highlights a key challenge in estimating sensitivity bounds --- as the gap between the bounding sensitivity functions widens, the estimand increasingly depends on tail properties of the outcome distribution, making nonparametric estimation more difficult and data-intensive.

We also provide a novel geometric connection between the Rosenbaum and Marginal models, relating to the diameter and radius of a set of density ratios. Moreover, we explore the practical implications of using one model over the other --- the Rosenbaum model is more suited to global sensitivity statements, while the Marginal model provides a targeted approach in settings where one has auxillary information or substantive domain knowledge about the nature of the unmeasured confounder. 



Overall, we hope that this work contributes toward a more flexible and interpretable toolkit for sensitivity analysis in the continuous treatment setting, and provides a foundation for future methodological and applied developments in this domain.


\bibliography{paper.bib}

\newpage
\appendix 

\section{Proof of Results in Section \ref{setup} and choices for sensitivity functions \texorpdfstring{$\Gamma_{t,t'}$}{Gamma(t,t')}} \label{setup-proof}

\subsection{Proof of Proposition \ref{expfamily}}
For (i), let $S(t) = \log \Upsilon_t$, then $\Gamma_{t,t'}$ can be written as $$\Gamma_{t,t'} = \exp(|S(t) - S(t')|).$$
Evidently, it is enough to produce a conditional distribution of $T|X,U$ to satisfy Assumption \ref{rosen-model}. Consider an exponential family for $T|X,U$ with sufficient statistic $S$, given by
$$f_{T|U,X}(t|X,U) = h(t)\exp\left[(\kappa(X) + \eta(X,U))S(t) - \lambda(X,U)\right],$$ where $\eta:\mathcal{X}\times \mathcal{U}\to [0,1]$.
Then, 
$$\dfrac{f_{T|U,X}(t|X,u)}{f_{T|U,X}(t|X,\tilde u)} = \exp\left[S(t)(\eta(X,u)-\eta(X,\tilde u))\right]\exp\left[\lambda(X,u) - \lambda(X,\tilde u)\right],$$
which yields 
\begin{align*}\dfrac{f_{T|U,X}(t|X,u)/f_{T|U,X}(t|X,\tilde u)}{f_{T|U,X}(t'|X,u)/f_{T|U,X}(t'|X,\tilde u)} &= \exp\left[(S(t) - S(t'))(\eta(X,u)-\eta(X,\tilde u))\right]\\ & \hspace{1em}\in [\exp(-|S(t)-S(t')|), \exp(|S(t)-S(t')|)],
\end{align*} thus completing the proof. 

The proof of (ii) follows from Proposition \ref{expfamily}(i) and Corollary \ref{corol1}(i). \hfill $\blacksquare$


\subsection{Choices for \texorpdfstring{$\Gamma_{t,t'}$}{Gamma(t,t')}}


\begin{longtable}{p{20em}l}
    \hline 
         \textbf{Distribution of $T|U$} & \textbf{Choice of Sensitivity Function}   \\ \hline \hline
         Normal($\gamma U$, 1) & $\exp(\gamma|t-t'|)$\\ 
         Normal(0,($\alpha + \gamma U)^{-1})$ & $\exp(\frac{\gamma}{2}|t^2-t'^2|)$\\ 
         Laplace($\gamma U$,1) & $\exp(2\gamma\mathbbm{1}[(t-\gamma)(t'-\gamma)<0])$\\
         Laplace(0, $(\alpha+\gamma U)^{-1})$ & $\exp(\gamma||t|-|t'||)$\\
         Gamma ($\alpha, \gamma U$) & $\exp(\gamma|t-t'|)$\\
         Gamma ($\gamma U, \lambda$) & $\exp(\gamma|\log t-\log t'|)$\\
         Beta $(\gamma\alpha(U), \gamma\beta(U))$ \hfill $\alpha(\cdot),\beta(\cdot)\in [0,1]$ & 
         $\left[\dfrac{(t\vee t')(1-t\wedge t')}{(t\wedge t')(1-t\vee t')}\right]^\gamma$
         \\
         Pareto (shape = $\gamma U$, rate = $\lambda$)  & $\exp(\gamma|\log t - \log t'|)$\\
         Weibull (rate = $\gamma U$, shape = $\alpha$) & $\exp(\gamma^\alpha |t^\alpha-t'^\alpha|)$\\
         Log-normal($\gamma U, 1)$ & $\exp(\gamma|\log t - \log t'|)$ \\
         Inverse Gamma (shape = $\alpha$, scale = $\theta + \gamma U$) & $\exp(\gamma|1/t - 1/t'|)$\\
         Logistic distribution($\gamma U$, 1) & $\exp(\gamma|t-t'|)\left[\dfrac{1+e^{-t\wedge t'}}{1+e^{-t\vee t'}}\cdot \dfrac{1+e^{-(t\vee t'-\gamma)}}{1+e^{-(t\wedge t'-\gamma)}}\right]^2$ \\
         \hline 
    \label{Sens-Function_tab}
\end{longtable}


\section{Definition of smooth function classes} \label{fnclass}
Let $k = (k_1,\cdots,k_d)$ be an index set, where each $k_i$ is non-negative integer, and let $|k| = \sum_{i=1}^d k_i$. For any $g:\T\to \R$, where $\T\subseteq \R^d$ and $g$ is differentiable upto $|k|$, let the differential operator $D^k$ be defined as $$D^kg = \dfrac{\partial^{|k|}g}{\partial^{k_1}t_1\cdots\partial^{k_d}t_d} \text{ and }D^0g = g.$$
For $s,L>0$, the H\"older class $\Sigma^H_d(s,L)$ on $\T$ is defined as 
$$\Sigma^H_d(s,L)=\left\{f:\T\to\R: \sum_{0\le |m|\le \lfloor s\rfloor} \|D^mf\|_{\infty} + \sum_{|k|=\lfloor s\rfloor}\sup_{t,t'\in\T, t\ne t'}\dfrac{|D^k f(t) - D^k f(t')|}{\|t-t\|^{s-|k|}}\le L\right\}.$$
The $L^p$-Sobolev space of order $m\in\N$, where $1\le p<\infty$, is defined by 
$$\Sigma^S_d(p,m)=\left\{f\in L^p: D^jf\in L^p\ \forall 1\le j\le m, \|f\|_p + \|D^mf\|_p<\infty\right\}.$$

\section{Bonvini-Kennedy-Ventura-Wasserman (BKVW) model on Bounding Density Ratios} \label{bkvw}


Some recent works in the continuous exposure literature have considered bounding the density ratio directly, rather than the ratio of density ratios. One such model was introduced in \cite{bonvini2022sensitivity} and \cite{jesson2022scalable} and is given by the following:


\textbf{BKVW model:} The law of $(T,X,U)$ satisfies 
\begin{equation}\dfrac{1}{\Delta}\le \dfrac{f_{T|U,X}(t|u,X)}{f_{T|X}(t|X)}\le \Delta \ \ \forall t\in\T, u\in\mathcal{U}.
\label{bkvw-eq}
\end{equation}
Although adopting a relatively simple form, the BKVW model contains subtle interpretational challenges that need to be addressed. First, note that the sensitivity bound $\Delta$ in Equation \eqref{bkvw-eq} is assumed to be uniform over all $t\in\mathcal{T}$. This uniformity can be difficult to satisfy, unless the exposure space $\mathcal{T}$ is compact.  While compactness is often a benign assumption in non-parametric inference, its explicit imposition in a sensitivity model complicates interpretation: the value of $\Delta$ then becomes dependent on how the exposure space is truncated, which can obscure its meaning. 

A potential solution one might consider is to allow the sensitivity bound to depend on $t$, ie, 
$$\dfrac{1}{\Delta_t} \le \dfrac{f_{T|U,X}(t|u,X)}{f_{T|X}(t|X)}\le \Delta_t \ \forall t\in \T, u\in\mathcal{U}, \ \ \text{a.s. } X,$$
a deeper issue remains: the interpretation of the sensitivity parameter becomes entangled with the observed distribution of the exposure $T$. To illustrate this, consider the case of no observed covariates, and a joint distribution of $$f_{T,U}(t,u) = f(t,u)g(u)f_T(t)/\delta_T(t),$$ 
where $\delta_T(t)$ is chosen such that $f_{T,U}$ is compatible with the observed distribution $f_T$ of $T$. 
Under such a model, we obtain, 
$$\dfrac{f_{T|U}(t|u)}{f_T(t)} = \dfrac{f(t,u)/\delta_T(t)}{\E[f(T,u)/\delta_T(T)]}$$
Assuming $U$ is compactly supported --- a standard assumption in sensitivity analyses --- the right hand side is uniformly bounded in $u$. However, this implies that the sensitivity function in the BKVW model depends not only on the strength of confounding through $f(t,u)$, but also on the observed distribution of the exposure $T$ via the denominator $\E[f(T,u)/\delta_T(t)]$, where $T\sim f_T(\cdot)$. In particular, the interpretation of the sensitivity parameter is no longer solely about the unmeasured confounding structure—it also hinges on the behavior of the marginal distribution of $T$. This dual dependence complicates the interpretability of the BKVW model in practice.

To see this in action, consider a treatment $T^*$, and $T$ to be a truncation of $T^*$ to the region $[-B,B]$. Such truncations are often common to remove outliers, or as technical non-parametric regression assumptions, as in Assumption \ref{tecn}, and are generally considered benign. However, the interpretation of the BKVW model becomes tricky under such truncations. To consider a concrete example, suppose $T^*|U=u$ has an exponential distribution with rate $\gamma U$, and $U\sim [0,1]$. Let $T$  be the truncation of $T^*$ to $[0,B]$ (ie, we discard any $T^*$ larger than $B$). In that case we obtain $$\dfrac{f_{T|U}(t|u)}{f_T(t)} = \left(\dfrac{1-\frac{1-e^{-\gamma B}}{\gamma B}}{1-e^{-\gamma u B}}\right)\left(\dfrac{\gamma^2t^2ue^{-\gamma u t}}{1-(\gamma t + 1) e^{-\gamma t}}\right).$$ 
Analogous expressions appear if $T^*|U\sim N(\gamma U,1)$ truncated at $\pm B$, in which case we obtain, 
$$\dfrac{f_{T|U}(t|u)}{f_T(t)} = \dfrac{\E[\Phi(B-\gamma U) - \Phi(-B-\gamma U)]}{\Phi(B-\gamma u) - \Phi(-B-\gamma u)}\cdot \dfrac{\gamma\phi(t-\gamma u)}{\Phi(t) - \Phi(t-\gamma)}.$$
The first terms here depend non-monotonically on $B$, and thus, even bounding the right hand side with $u\in [0,1]$ results in a $\Delta_t$ that also depends on the truncation parameter $B$. This complicates the interpretation of the sensitivity function, which now also intrinsically depends on the arbitrary truncation point $B$. 




\section{Competing Models Results: Proof of Results in Section \ref{comparison}}
\subsection{Proof of Proposition \ref{geometry}} \label{geom-proof}
(i) follows trivially based on Equation \eqref{rosen-eq} and the definition of $S_{t,t'|X}$. 
    
For (ii), We first claim that $g_{t,t'|X}$ admits an intermediate value property (IVP), ie, for any $a, b\in\mathcal{U}$, if $y$ is a real value between $g_{t,t'|X}(a)$ and $g_{t,t'|X}(b)$, then $\exists c\in\mathcal{U}$ such that $g_{t,t'|X}(c) = 0$.

To see this, assume wlog that $g_{t,t'|X}(a)<y<g_{t,t'|X}(b)$, and consider the sets $A = g_{t,t'|X}^{-1}(\infty,y)$ and $B = g_{t,t'|X}^{-1}(y,\infty)$. Both of these are open in $\mathcal{U}$ since $g_{t,t'|X}$ is continuous. Moreover both of these are non-empty as $a\in A$ and $b\in B$, and are clearly disjoint. Thus, if $\not\exists c: g_{t,t'|X}(c) = y$, then $A\cup B = \mathcal{U}$, contradicting that $\mathcal{U}$ was connected. Hence the intermediate value property must hold. 

Now, given that IVP holds for $g_{t,t'|X}$, we claim that $f_{T|X}(t|X)/f_{T|X}(t'|X)$ belongs to the range of $g_{t,t'|X}$. To prove this, suppose the contrary. Then by IVP, either $g_{t,t'|X} <f_{T|X}(t|X)/f_{T|X}(t'|X) \ \ \forall u\in\mathcal{U}$, or the reverse inequality must hold for all $u\in\mathcal{U}$. Wlog assume the former. Then, by definition of $g_{t,t'|X}$, we have,
$$\dfrac{f_{T|U,X}(t|u,X)}{f_{T|U,X}(t'|u,X)}<\dfrac{f_{T|X}(t|X)}{f_{T|X}(t'|X)} \ \forall u\in\mathcal{U}\implies f_{U|T,X}(u|t,X)<f_{U|T,X}(u|t',X) \ \forall u,$$
by Bayes theorem. Thus $f_{U|T,X}(u|t',X) - f_{U|T,X}(u|t,X)$ is a strictly positive function, whose integral on $\mathcal{U}$ constitutes to $0$ (since both of them are valid densities and hence integrate to 1 over $\mathcal{U}$), which is a contradiction. Hence, there exists $u^*\in\mathcal{U}$ such that $g_{t,t'|X}(u^*) = f_{T|X}(t|X)/f_{T|X}(t'|X)$. 

(ii) thus follows by equating $f_{T|X}(t|X)/f_{T|X}(t'|X)$ to $g_{t,t'|X}(u^*)$ and taking logarithms. \hfill $\blacksquare$

\subsection{Proof of Corollary \ref{corol1}} \label{corol1-proof}
Let $r$ be the radius of a set $\mathcal{U}$ from a point $u^*$ wrt to a metric $d$, and let $D$ be the diameter of $\mathcal{U}$. We begin by proving 
$$r\le D\le 2r.$$
For any point $u\in \mathcal{U}$, $d(u,u^*)\le D\implies r = \sup_{u} d(u,u^*)\le D$, proving the first inequality. For the second inequality, for any two points $u,v\in\mathcal{U}$, $d(u,v)\le d(u,u^*)+d(v,u^*)\le 2r$ by triangle inequality. The inequality follows by taking supremum over all possible $u,v\in\mathcal{U}$. 

From Proposition \ref{geometry} we already know that the Rosenbaum model defines the diameter of the set $S_{t,t'|X}$ as $\log\Gamma_{t,t'}$, while the Marginal model defines its radius from a point $u^*\in\mathcal{U}$ (when $\mathcal{U}$ is connected and $g_{t,t'|X}$ is continuous, or more generally from $f_{T|X}(t|X)/f_{T|X}(t'|X)$) by $\log\Lambda_{t,t'}$. Corollary \ref{corol1} thus follows from the above inequality. \hfill $\square$

\subsection{Proof of Theorem \ref{data-compatibility}} \label{data-compatibility-proof}

Throughout this proof, we shall keep the conditioning on $X$ implicit, so that all following constructions may be considered implicit on strata of $X$. 

Proof of (1):

\begin{lemma}
\label{uni-exi}
    Consider a typical element of $\F$, characterized by $\Upsilon_t\ge 0$. Then, $\exists$ a joint distribution of $(T,U)$ such that $T,U\sim U[0,1]$, and $f_{T,U}(t,u) = \exp(S(t)\eta(u))h(t)g(u)$, where $S(t) = \log \Upsilon_t$, and $\eta(u)\in[0,1]$.
\end{lemma}

\textit{Proof}: We shall use the following result from \cite{borwein1994entropy}. 

\begin{result}[\cite{borwein1994entropy}, Corollary 3.6]
\label{borwein}
    Let $\phi:\R\to(-\infty,\infty]$ be 
    $\phi(r) = r\log(r/e)\mathbbm{1}(r>0) + \infty \cdot \mathbbm{1}(r<0)$ (interpreting $\infty\cdot 0 = 0$), 
    and consider the optimization problem 
    \begin{align*}
        \inf &\int_{\mathcal{U}\times \T} \phi(v(u,t))k(u,t)\,du\,dt & \text{subject to} &\int_{\T} v(u,t)k(u,t) \,dt = \alpha(u) \text{ a.e. on }\mathcal{U}\\
         v \in & L_1(\mathcal{U}\times \mathcal{T}, du\times dt) & & \int_{\mathcal{U}} v(u,t)k(u,t)\,du = \beta(t) \text{ a.e. on }\T
    \end{align*}
    where $\mathcal{U}$ and $\T$ are finite measure spaces. If $\exists \tilde v>0$ a.e. $[k \,du \,dt]$ feasible for the above optimization problem, then $\exists$ a unique optimizing solution $v_0$ of the above, and $\exists h_0:\T\to (0,\infty)$ and $g_0:\mathcal{U}\to (0,\infty)$ such that $v_0(u,t) = h_0(t)g_0(u)$.
\end{result}

For lemma \ref{uni-exi}, $\mathcal{U}=\T=[0,1]$, and $\alpha(u)=1$ a.e. on $[0,1]$, and $\beta(t) = 1$ a.e. on $[0,1]$. Moreover, take $k(u,t) = \exp(S(t)\eta(u))$. Then, one feasible solution is $\tilde v(u,t) = \exp(-S(t)\eta(u))$. Hence, by Result \ref{borwein}, $f_{T,U} = v_0$, the unique optimizer to the minimization problem. \hfill $\square$

Lemma \ref{uni-exi} proves that a $(T,U)$ can be constructed to be marginally uniform, and jointly of the specified form, and hence satisfying Assumption \ref{rosen-model}. For any arbitrary marginals $F_T$ and $F_U$, let $(T',U')$ be constructed using Lemma \ref{uni-exi}, such that they are marginally uniform, and jointly given by $f_{T',U'}(t,u) = \exp(\delta(t)\eta(u))h(t)g(u)$, where $\delta(t)$ and $\eta(u)\in[0,1]$ are to be suitably chosen. 

Define $T = F_T^{-1}(T')$ and $U = F_U^{-1}(U')$. Let $F_T$ be absolutely continuous with respect to a base measure $\mu_T$ on $\T$ (this can always be done without loss of generality, since any measure is absolutely continuous with respect to itself). 

Then, 
$$\dfrac{d\Q_{T|U=u}}{d\mu_T}(t) = f_{T',U'}(F_T(t),F_U(u))\,\dfrac{dF_T}{d\mu_T}(t) = \exp(\delta(F_T(t))\eta(F_U(u))) h(F_T(t))g(F_U(u))\dfrac{dF_T}{d\mu_T}(t)$$
Hence, 
\begin{align}
    \dfrac{\dfrac{d\Q_{T|U=u}}{d\mu_T}(t)/ \dfrac{d\Q_{T|U=\tilde u}}{d\mu_T}(t)}{\dfrac{d\Q_{T|U=u}}{d\mu_T}(t')/ \dfrac{d\Q_{T|U=\tilde u}}{d\mu_T}(t')} &= \exp\left((\delta(F_T(t)) - \delta(F_T(t')))(\eta(F_U(u))-\eta(F_U(u')))\right) \nonumber \\
    &\in \exp(\pm |\delta(F_T(t))-\delta(F_T(t'))|) \label{inter-step}
\end{align}

Let $\Gamma_{t,t'}$ be generated by a typical element $\Upsilon_t$ such that $\Gamma_{t,t'}\in \F$. Since $\Gamma_{t,t'}$ is measurable with respect to $F_T\times F_T$, and $\Upsilon_t$ is not identically 0, $\Upsilon_t$, and hence $S(t) := \log \Upsilon_t$ are measurable with respect to the observed measure $F_T$. 
Moreover, as in Appendix \ref{setup-proof}, $\Gamma_{t,t'} = \exp(|S(t)-S(t')|)$. 

Since $S$ is $F_T$-measurable, we can use Doob-Dynkin factorization lemma (\cite{kallenberg1997foundations}, Lemma 1.14), to decompose $S = \delta^* \circ F_T$, where $\delta^* \in\mathcal{B}([0,1])$. Choosing $\delta(t) = \delta^*(t)$ allows us to conclude from Equation \eqref{inter-step} that $(T,U)$ satisfies Assumption \ref{rosen-model}.

Furthermore, if $\eta$ is chosen to be such that $\eta(u)\to 1$ as $u\to \infty$ and $0$ as $u\to -\infty$, then Equation \eqref{inter-step} can be strengthened to obtain 
$$\sup_{u,u'\in\mathcal{U}} \dfrac{\dfrac{d\Q_{T|U=u}}{d\mu_T}(t)/ \dfrac{d\Q_{T|U=\tilde u}}{d\mu_T}(t)}{\dfrac{d\Q_{T|U=u}}{d\mu_T}(t')/ \dfrac{d\Q_{T|U=\tilde u}}{d\mu_T}(t')} = \Gamma_{t,t'}.$$

Finally, to construct $Y(t)$, we need to ensure the distribution of the observable $Y|T$ matches that of the observed $F_{Y|T}$. Let $$Y' = F_{U|T}^{-}(U) + V\cdot\Delta_{F_{U|T}}(U),$$ 
where $F_{U|T}$ is the conditional CDF of $U|T$ based on the constructed $\Q$, $$\Delta_{F_{U|T}}(u):= F_{U|T}(u) - F_{U|T}^-(u),$$ and $V\sim U[0,1]$, $V\indep U, T$. since the cdf of $Y'$ mimics that of $F_{U|T}(U)$ at all continuity points, and linearly interpolates between $F_{U|T}^-$ and $F_{U|T}$ at the discontinuity points, it is easy to check that $Y'|T \sim U[0,1]$. Define $Y(t) = F_{Y|T}(Y'|t)$ for all $t\in\mathcal{T}$. Thus, $Y(T) = Y|T\sim F_{Y|T}$. 

Moreover, since $\{Y(t)\}_{t\in\mathcal{T}}$ are measurable with respect to $(U,V)$, and $V$ is independent of $T$, we have $\{Y(t)\}_{t\in\mathcal{T}} \indep T|U$ as required. This concludes the proof of (1). \hfill $\square$

Proof of (2):

By Bayes theorem, the Marginal model is equivalent to $$\dfrac{1}{\Lambda_{t,t'}} \le \dfrac{f_{U|T}(u|t)}{f_{U|T}(u|t')}\le \Lambda_{t,t'}.$$

Thus, for construction of full-data random variables compatible with the observed data, one can only construct $U|T$ appropriately. The construction of $\{Y(t)\}_{t\in\mathcal{T}}$ is similar to that in (1), and is hence omitted. 

Let $S(t) = \dfrac 12\log \Upsilon_t$, and thus $\Lambda_{t,t'} = \exp(2|S(t)-S(t')|)$, and consider the exponential family
$$f_{U|T}(u|t) = \exp(\eta(u)S(t))h(t)g(u),$$ where $\eta(u)\in [-1,1]$ $$h(t) = \left(\int_{\mathcal{U}} \exp(\eta(u)S(t))g(u) \,du\right)^{-1}$$ to ensure $f_{U|T}$ is a valid density. 

Now,
\begin{align*}
\dfrac{1}{h(t')} &= \int_{\mathcal{U}} \exp(\eta(u)S(t'))g(u)\,du = \int_{\mathcal{U}} \exp(\eta(u)(S(t') - S(t)))\cdot \exp(\eta(u)S(t))g(u)\,du\\ 
&\le \exp(|S(t)-S(t')|)\int_{\mathcal{U}} \exp(\eta(u)S(t))g(u)\,du = \dfrac{\exp(|S(t) - S(t')|)}{h(t)}.
\end{align*}

Thus,
$$\dfrac{f_{U|T}(u|t)}{f_{U|T'}(u|t')} = \exp(\eta(u)(S(t)-S(t'))\dfrac{h(t)}{h(t')}\le \exp(2|S(t)-S(t')|) = \Lambda_{t,t'}.$$ 
The opposite inequality with $\Lambda_{t,t'}^{-1}$ follows analogously. This completes the proof. \hfill $\blacksquare$

\subsection{Proof of Remark \ref{rem:mar-nonglobal}} \label{mar-nonglobal-proof}

Consider $T$ and $U$ to be dichotomous on $\{0,1\}$, with fixed marginals. 
If we denote $p_{00} = p$ as the free parameter, then, the $2\times 2$ table is as in Table \ref{tab:2*2} 
\begin{table}[!ht]
    \centering
    \begin{tabular}{|c|cc|c|} \hline 
         & $U=0$ & $U=1$ & \\ \hline 
         $T=0$ & $p$ & $1-p_T - p$ & $1-p_T$\\
         $T=1$ & $1-p_{U} - p$ & $p_U+p_T -1 +p $ & $p_T$\\ \hline
         & $1-p_U$ & $p_U$ & 1\\ \hline 
    \end{tabular}
    \caption{$T\times U$ dichotomous distribution.}
    \label{tab:2*2}
\end{table}

Then, 
\begin{align*}
    \dfrac{\P(T=1|U=1)/\P(T=0|U=1)}{\P(T=1)/\P(T=0)} &= \dfrac{1-p_T}{p_T}\cdot \dfrac{p_T+p_U-1+p}{1-p_T-p} &=:O_1(p)\\
    \dfrac{\P(T=1|U=0)/\P(T=0|U=0)}{\P(T=1)/\P(T=0)} &= \dfrac{1-p_T}{p_T}\cdot \dfrac{1-p_U-p}{p} &=:O_2(p)
\end{align*}

And thus, 
$$\max_p \max(O_1(p),O_2(p)) = \begin{cases} \dfrac{1-p_T}{p_U - p_T} & \text{ if } p_T<p_U \ \& \ p_T+p_U >1 \\ \infty &\text{ otherwise}\end{cases}$$
Thus, choosing $p_T$ and $p_U$ so that the first constraint is satisfied, makes the range of $\Lambda_{1,0}$ (the only relevant bounding quantity for binary $T$), or equivalently the $\max_p \max(O_1(p), O_2(p))$, restricted, and hence the supremum cannot vary freely in $\R_{\ge 0}$. \hfill $\blacksquare$

\section{Identification Results: Proof of Results in Section \ref{partial-id}}
\begin{lemma}
\label{lem:restr}
    Let the full-data distribution satisfy latent ignorability (Assumption \ref{uc}) and (i) either Rosenbaum sensitivity model (Assumption \ref{rosen-model}), (ii) or Marginal sensitivity model (Assumption \ref{mar-model}), and let $L_{t,t'}$ define the likelihood ratio:
    $$L_{t,t'}(y,x) = \dfrac{d\P(Y(t)\in \cdot\ |T=t',X=x)}{d\P(Y(t)\in \cdot\ |T=t,X=x)}(y). $$
    Then $\P_{Y(t)|T=t',X=x}$ is absolutely continuous with respect to $\P_{Y(t)|T=t,X=x}$, and the likelihood ratio $L_{t,t'}$ satisfies 
    \begin{itemize}
    \item[(i)] $0\le L_{t,t'}(y,x)\le \Gamma_{t,t'}L_{t,t'}(\tilde y, x)$ under the Rosenbaum sensitivity model.
    \item[(ii)] $\Lambda_{t,t'}^{-1} \le L_{t,t'}(y,x)\le \Lambda_{t,t'}$, under the Marginal sensitivity model.
    \end{itemize} 
    Furthermore, if $\Gamma_{t,t'}$ (or $\Lambda_{t,t'}$) chosen from $\mathcal{F}$, and for any $L_{t,t'}(y,x)$ satisfying the inequality in (i)  (or (ii)), there is a distribution $P$ that satisfies Assumption \ref{uc} and \ref{rosen-model} (or \ref{mar-model}).
\end{lemma}

\textit{Proof}: 

\textbf{Step 1}: \textit{(Absolute continuity)}

Since $(T,U,X)$ satisfies one of the sensitivity models in Assumption \ref{rosen-model} or \ref{mar-model}, the regular conditional distribution $\P_{T|U=u,X=x}(\cdot), u\in\mathcal{U}, x\in\mathcal{X}$ is absolutely continuous wrt some base measure $\mu$ for $\P_{U,X}$ almost everywhere $u,x$. 

Thus, $T|X=x$ is also absolutely continuous wrt $\mu$ for $\P_X$ almost everywhere $x$, and admits a valid density $f_{T|X}(\cdot|x)$ wrt $\mu$. 

Now, for any measurable set $A\subset \mathcal{U}$,
\begin{align*}\dfrac{\P(U\in A|T=t',X=x)}{\P(U\in A|T=t,X=x)} &= 
\dfrac{\int_A d\P_{U|T=t', X=x}(u)}{\int_A d\P_{U|T=t, X=x}(u)} \\
&= \dfrac{\int_A f_{T|U,X}(t'|u',x)\,d\P_{U|X=x}(u')}{\int_A f_{T|U,X}(t|u,x)\,d\P_{U|X=x}(u)}\cdot \dfrac{f_{T|X}(t|x)}{f_{T|X}(t'|x)}
\end{align*}
If the Rosenbaum model holds, then by the quasi-convexity of the ratio mapping $(a,b)\mapsto a/b$, the above ratio lies within $[\Gamma_{t,t'}^{-1},\Gamma_{t,t'}]$. On the other hand, if the Marginal model holds, the same ratio $\in [\Lambda_{t,t'}^{-1},\Lambda_{t,t'}]$.

Denoting $q_t(u|x)$ as the density of $U|T=t,X=x$ (wrt some base measure $\mu$) we thus have $r_{t,t'}(u,x) = \dfrac{q_{t'}(u|x)}{q_t(u|x)}\in \left[\Gamma_{t,t'}^{-1}, \Gamma_{t,t'}\right]$ under (i), or in $\left[\Lambda_{t,t'}^{-1},\Lambda_{t,t'}\right]$ under (ii). Hence, by latent ignorability, for any measurable set $A\subset\R$, 
\begin{align*}
    \dfrac{\P(Y(t)\in A|T=t',X=x)}{\P(Y(t)\in A|T=t, X=x)} &= \dfrac{\int \P(Y(t)\in A|T=t',X=x, U=u) q_{t'}(u|x)\, d\mu(u)}{\int \P(Y(t)\in A|T=t, X, U=u) q_t(u|x)\, d\mu(u)}\\
    &= \dfrac{\int \P(Y(t)\in A|X=x, U=u) q_{t'}(u|x)\, d\mu(u)}{\int \P(Y(t)\in A|X=x, U=u) q_t(u|x)\, d\mu(u)} \\
    &\in \begin{cases} [\Gamma_{t,t'}^{-1}, \Gamma_{t,t'}] &\text{ under (i)} \\ [\Lambda_{t,t'}^{-1},\Lambda_{t,t'}] &\text{ under (ii)} \end{cases}.
\end{align*}
This proves the absolute-continuity. 

\textbf{Step 2}: \textit{(Likelihood ratio as a conditional expectation)}

Denoting $r_{t,t'}(u,x) = q_{t'}(u|x)/q_t(u|x)$ as in Step 1, note that for any set $B$ measurable with respect to the sigma-algebra of $Y(t)$, we have
$$\E[\1(B)|T=t', X=x] = \E[\1(B)\E[r_{t,t'}(U)|Y(t), T=t, X=x]|T=t, X=x]$$
holding almost surely, thus implying by the Radon-Nikodyn theorem that the likelihood ratio  $L_{t,t'}$ satisfying 
$$L_{t,t'}(y,x) = \E[r_{t,t'}(U,x)|Y(t) = y, T=t, X=x].$$

\textbf{Step 3}: \textit{Constraints on likelihood ratio under (i) and (ii)}

(i) Consider the Rosenbaum sensitivity model as in Assumption \ref{rosen-model}. Then,
\begin{equation} \dfrac{r_{t,t'}(u,x)}{r_{t,t'}(\tilde u,x)} = \dfrac{q_{t'}(u|x)}{q_t(u|x)}\cdot \dfrac{q_t(\tilde u|x)}{q_{t'}(\tilde u|x)} = \dfrac{f_{T|U,X}(t'|u,X=x)}{f_{T|U,X}(t|u,X=x)}\cdot \dfrac{f_{T|U,X}(t|\tilde u,x)}{f_{T|U,X}(t|\tilde u,x)}\le \Gamma_{t,t'}. \label{bayes}
\end{equation}
Fix an $\varepsilon>0$, and let $u_0$ be such that $r_{t,t'}(u_0,x)\le \inf_u r_{t,t'}(u,x) + \varepsilon$, then, by Equation \eqref{bayes},
$$L_{t,t'}(y,x) = r_{t,t'}(u_0,x)\E\left[\left.\dfrac{r_{t,t'}(U,x)}{r_{t,t'}(u_0,x)}\right|Y(t), T=t, X=x\right]\le \Gamma_{t,t'}r_{t,t'}(u_0,x).$$
On the other hand, 
$$L_{t,t'}(\tilde y,x)\ge \inf_u r_{t,t'}(u,x) \ge r_{t,t'}(u_0,x) -\varepsilon.$$
Thus, 
$$L_{t,t'}(y,x) \le \Gamma_{t,t'} r_{t,t'}(u_0,x) \le \Gamma_{t,t'}(L_{t,t'}(\tilde y,x) + \varepsilon),$$
and the proof of (i) is complete by noticing that $\varepsilon >0$ is arbitrary.

(ii) For the Marginal sensitivity model in Assumption \ref{mar-model}, note that 
\begin{equation} \label{mar-bayes}
r_{t,t'}(u,x) = \dfrac{q_{t'}(u|x)}{q_t(u|x)} \in [\Lambda_{t,t'}^{-1},\Lambda_{t,t'}].
\end{equation}
(ii) Thus follows trivially by Equation \ref{mar-bayes} and Step 2.

The converse inclusion is trivial, by taking $U = \{Y(t)\}_{t\in \mathcal{T}}$ to trivially satisfy latent ignorability, and taking the density of $f_{T|U,X}(t|U=u, X=x)$ to be constructed as in Appendix \ref{setup-proof}. \hfill $\blacksquare$



\subsection{Proof of Theorem \ref{id-rosen}} \label{id-rosen-proof}

We present the proof keeping $X=x$ implicit. Let $\mathcal{L}_{t,t'}$ denote the space of all non-negative real valued measurable functions $L_{t,t'}$ such that $L_{t,t'}(y)\le \Gamma_{t,t'}L_{t,t'}(\tilde y)$ for all $y$, $\tilde y$. By Lemma \ref{lem:restr}(i), it suffices to search for $L_{t,t'}\in\mathcal{L}_{t,t'}$ for the purposes of Theorem \ref{id-rosen} . Clearly $\mathcal{L}_{t,t'}$ is convex, non-empty (since $1\in \mathcal{L}_{t,t'})$, and for $L_{t,t'}\equiv 1$, $\E[L_{t,t'}(Y(t))|T=t] = 1$. Thus, by strong duality theorem (\cite{luenverger1969optimization}, Theorem 8.6.1), we have, 
\begin{align*} & \inf_{L\in \mathcal{L}_{t,t'}}\{\E[Y(t)L_{t,t'}(Y(t))|T=t] : \E[L_{t,t'}(Y(t))|T=t] = 1\}  \\ &= \sup_{\mu\in \R} \inf_{L_{t,t'}\in \mathcal{L}_{\Gamma_{t,t'}}} \{\E[(Y(t) - \mu)L_{t,t'}(Y(t))|T=t] + \mu\}.
\end{align*}
Now, note that for this optimization problem, since $L_{t,t'}\in \mathcal{L}_{t,t'}$, hence $L(y) \le \Gamma_{t,t'} L(\tilde y)$ for all $y,\tilde y$. Fixing $y$ and taking $\tilde y$ arbitrary allowed us to infer $L_{t,t'}(y) \le \Gamma_{t,t'} \inf_{\tilde y} L_{t,t'}(\tilde y)$, and hence $c := \inf_{\tilde y} L_{t,t'}(\tilde y)$ is non-negative finite. Thus, $L_{t,t'}(y)\le \Gamma_{t,t'} c$ for all $y$, and hence $L_{t,t'}(y) \in [c,c\Gamma_{t,t'}]$ for all $y$. 

Now we claim that a minimizer $L^*_{t,t'}$ of the above optimization problem can be found by restricting oneself to take values in $\{c, c\Gamma_{t,t'}\}$ only. To see this, note that for any $L_{t,t'} \in [c,c\Gamma_{t,t'}]$, one may construct a function $L^*_{t,t'}(y) = c\mathbbm{1}(y-\mu >0) + c\Gamma_{t,t'}\mathbbm{1}(y-\mu\le 0)$, such that $(y-\mu)L_{t,t'}^*(y) \le (y-\mu)L_{t,t'}(y)$, with equality only if $y = \mu$. Thus, any function $L_{t,t'}\in \mathcal{L}_{t,t'}$ can be modified into the form
$$L_{t,t'}^*(y)\propto \Gamma_{t,t'}\1\{y-\mu\le 0\} + \1\{y-\mu >0\}$$
without increasing the objective function. 

Using the above form for $L_{t,t'}$, one thus obtains $$\theta_{t'}(t) = \sup_\mu \inf_{c\ge 0} \left\{\E[c\psi^{t,t'}_\mu(Y(t))|T=t] +\mu\right\}.$$
Define $R = \{\mu: \E[\psi^{t,t'}_\mu(Y(t))|T=t] \ge 0\}$, the above can be simplified as 
\begin{align*} \theta_{t'}(t) &= \max\left\{\sup_{\mu\in R}\inf_{c\ge 0} \left(\E[c\psi^{t,t'}_\mu(Y(t))|T=t] +\mu\right), \sup_{\mu\in R^c}\inf_{c\ge 0}\left(\E[c\psi^{t,t'}_\mu(Y(t))|T=t] +\mu\right)\right\} \\
&= \max\left\{\sup_{\mu\in R} \mu, \sup_{\mu\in R^c} -\infty\right\} = \sup (R).
\end{align*}
Next, note that $\mu\mapsto \E[\psi_\mu^{t,t'}(Y(t))|T=t]$ is a nonincreasing function with derivative, $-[(\Gamma_{t,t'} - 1)\P(Y(t)\le \mu|T=t)+1]$. Next, note that since $|\theta_{t'}(x,t)|<\infty$, $\sup(R)$ is finite; and $\sup(R)$ must lie in the convex hull of the support of $Y(t)|T=t$, as $\psi_\mu^{t,t'}(Y(t))$ is non-positive (non-negative) at the supremum (infimum) of the convex hull of the support of $Y(t)|T=t$. Since $Y(t)|T=t$ assumes a positive density on the convex hull of it's support, the function $\P(Y(t)\le \mu|T=t)$ is non-constant for all $\mu$ in the convex hull of the support, and hence the mapping $\mu\mapsto \E[\psi_\mu^{t,t'}(Y(t))|T=t]$ is strictly increasing in the convex support, allowing $\sup(R)$ to be written as the unique root of the equation $\E[\psi_\mu^{t,t'}(Y(t))|T=t]$ as required. 

Finally, if $\E[(Y-\theta_{t'}(X,t))^2_+ + \Gamma_{t,t'}(Y-\theta_{t'}(X,t))^2_-|T=t,X]<\infty$, then by Normal integrand theory (\cite{rockafellar2009variational}, Theorem 14.60), $\theta_{t'}(X,t)$ is a normal integrand, and the provided optimization problem is convex, allowing $\theta_{t'}(X,t)$ to be the almost surely unique minimizer of the provided optimization problem. \hfill 
$\blacksquare$


\subsection{Proof of Theorem \ref{id-mar}} \label{id-mar-proof}


As in Appendix \ref{id-rosen-proof}, we present the proof keeping $X=x$ conditioning implicit. Let $\mathcal{M}_{t,t'}$ denote the space of all non-negative real valued measurable functions $L_{t,t'}$ such that $\Lambda_{t,t'}^{-1}\le L_{t,t'}(y)\le \Lambda_{t,t'}$, which is sufficient to consider based on Lemma \ref{lem:restr}. 
%
Hence we need to obtain $$\zeta_{t'}(t) = \inf_{L\in \mathcal{M}_{t,t'}}\{\E[Y(t)L_{t,t'}(Y(t))|T=t]: \E[L_{t,t'}(Y(t))|T=t]\}.$$
Again, since $\mathcal{M}_{t,t'}$ is convex, non-empty (since it contains the constant 1 function), and has a non-empty intersection with the feasibility restriction $\E[L_{t,t'}(Y(t))|T=t]=1$ again via the constant 1 function, we apply strong duality to obtain:
$$\zeta_{t,t'}(y) = \sup_{\mu\in \R} \inf_{L\in \mathcal{M}_{t,t'}} \{\E[(Y-\mu)L(Y)|T=t] +\mu\}.$$
Next, for any $\mu\in \R$, note that the optimizer $L^*_{t,t'}(Y)$ can be chosen to take values $\Lambda_{t,t'}$ when $Y\le \mu$ and $\Lambda_{t,t'}^{-1}$ when $Y>\mu$, since any other function $L_{t,t'}(Y)\in \mathcal{M}_{t,t'}$ can be modified into the above form without increasing the objective function. This yields, 
$$\zeta_{t,t'}(y) = \sup_{\mu\in \R}\{\mu + \E[\rho^{t,t'}_\mu(Y)|T=t]\}.$$
Now, using $\mu =\E[Y|T=t] - \E[Y-\mu|T=t] = \E_t[Y] - \E_t[(Y-\mu)_+] + \E_t[(Y-\mu)_-]$, we obtain, 
\begin{align*} \mu + \E[\rho^{t,t'}_\mu(Y)|T=t]\} &= \E[Y|T=t] - \E\left[\left.(1-\Lambda_{t,t'}^{-1})(Y-\mu)_+ +(\Lambda_{t,t'}-1)(Y-\mu)_-\right|T=t\right] \\
&= \E[Y|T=t] -\left(\Lambda_{t,t'}-\Lambda_{t,t'}^{-1}\right)\E\left[\left.\dfrac{1}{\Lambda_{t,t'}+1}(Y-\mu)_+ + \dfrac{\Lambda_{t,t'}}{\Lambda_{t,t'}+1}(Y-\mu)_-\right|T=t\right]
\end{align*}
Thus, maximizing this over $\mu$ is equivalent to minimizing the pinball-loss function, which occurs at the known optimum $Q_{Y|T=t}\left(\dfrac{1}{\Lambda_{t,t'}+1}\right).$ This completes the proof. \hfill $\blacksquare$

\section{Proof of Semi-parametrics Results in Section \ref{inflfn}}

For the purposes of proving Theorem \ref{rosen-eif} and \ref{mar-eif}, we will represent the entire observed data joint distribution with the CDF $F$. Moreover, $\psi_R$ and $\psi_M$ will be represented by a generic functional $\psi$, where $\psi$ shall represent $\psi_R$ in Subsection \ref{rosen-eif-proof} and $\psi_M$ in Subsection \ref{mar-eif-proof}. 

Consider a parametric sub-model $F(W;\eta)$ with parameter $\eta\in \R$, such that $F(W;\eta) = F(W;0)$. For exposition, one may consider $dF(W;\eta) = \exp(\eta b(W))dF(W)$ such that $b(W)$ admits a moment-generating function in a neighborhood of 0 under the law of $W$. 

Under a non-parametric model, the efficient influence function for $\psi$ is the unique mean-zero function $\phi(W)$ such that the pathwise derivative of $\psi$ under any parametric sub-model equals the inner-product of $\phi$ and the score-function of that parametric sub-model. In other words, \begin{equation} \label{inf-def}\left.\dfrac{d}{d\eta}\psi_\eta\right|_{\eta = 0} = \E[\phi(W)S(W;0)], \end{equation}
where $S(W;\eta) = \dfrac{\partial}{\partial\eta}\log dF(W;\eta)$.

Now, since the joint likelihood $dF(W;\eta)$ can be factorized as $dF(Y|T,X;\eta)dF(T|X;\eta)dF(X;\eta)$, this allows one to write 
\begin{equation} S(W;\eta) = S(Y|T,X;\eta) + S(T|X;\eta) + S(X;\eta). \label{score-split} \end{equation}
Next, note that if $$\psi = \int\int\int h_{t'}(x,t)dF(t'|x)dF(x)dF(t),$$
as is the case for both $\psi_R$ and $\psi_M$, then,
\begin{align} \label{der-expand}
    \left.\dfrac{d}{d\eta}\psi_\eta\right|_{\eta = 0} &= \left.\dfrac{d}{d\eta}\int \int\int h_{t'} (x,t;\eta)dF(t'|x;\eta)dF(x;\eta)dF(t;\eta)\right|_{\eta = 0} \nonumber\\
    &= \int \int\int\left.\dfrac{\partial}{\partial\eta}h_{t'}(x,t;\eta)\right|_{\eta = 0}dF(t'|x;0)dF(x;0)dF(t;0) \nonumber\\ & \hspace{2em} + 
    \int \int\int h_{t'}(x,t;0) (S(t'|x;0) + S(x;0) + S(t;0))dF(t'|x;0)dF(x;0)dF(t;0)
\end{align}

\begin{lemma} \label{sec-eif}
    The second term in Equation \eqref{der-expand} can be simplified as 
    $$\E[(A(T,X) + B(T))(S(T|X) + S(X))],$$
    where $A(t,x) = \E[h_t(x,T)]$ and $B(t) = \E[h_t(X,T)]$. Equivalently, the second term can also be written as 
    $$\E[(A(T,X) + B(T))S(W)]$$
\end{lemma}

\textit{Proof}: 
\begin{align}
    & \int\int\int h_{t'}(x,t;0)[S(t'|x;0) + S(x;0) + S(t;0)] dF(t'|x;0) dF(x;0)dF(t;0) \nonumber \\
    & \int\int\int\int h_{t'}(x,t;0)[S(t'|x;0) + S(x;0) + S(t|x';0) + S(x';0)] dF(t'|x;0)dF(x;0)dF(t|x;0)dF(x';0), \label{S-expand}
\end{align} 
as $$S(t;0)dF(t;0) = \int (S(t|x';0) + S(x';0))f(t|x';0)dF(x';0).$$
Continuing from Equation \eqref{S-expand},
\begin{align}
    & \int\int\int h_{t'}(x,t;0)[S(t'|x;0) + S(x;0) + S(t;0)] dF(t'|x;0) dF(x;0)dF(t;0) \nonumber \\
    &= \int\int \underbrace{\left[\int \int  h_{t'}(x,t;0) dF(t|x';0)dF(x';0)\right]}_{=:A(t',x)}(S(t'|x;0) + S(x;0)) \,dF(t'|x)dF(x) \nonumber \\ 
    & + \hspace{2em} \int\int \underbrace{\left[\int\int h_{t'}(x,t;0)\,dF(t'|x;0)dF(x;0)\right]}_{=:B(t)} (S(t|x';0) + S(x';0)) \,dF(t|x')\,dF(x') \nonumber \\
    &= \int\int A(t',x) (S(t'|x;0) + S(x;0))dF(t'|x;0)dF(x;0) \nonumber \\ &\hspace{2em} + \int\int B(t)(S(t|x';0) + S(x';0))dF(t|x';0)dF(x';0) \nonumber \\
    &= \int\int (A(t,x) + B(t)) (S(t|x;0) + S(x;0))\,dF(t|x;0)\,dF(x;0) \nonumber 
\end{align}
where the last step follows by unifying the dummy variables of integration. 

The second part of the lemma follows from the fact that $\E[S(Y|T,X)|T,X] = 0$. \hfill $\square$

\subsection{Proof of Theorem \ref{rosen-eif}} \label{rosen-eif-proof}

By Theorem \ref{data-compatibility}, the Rosenbaum model does not impose any restrictions on the observed data distribution. Thus the observed data tangent space under Assumption \ref{rosen-model} is a superset of the tangent space under a non-parametric model, which is known to be all of $L^2$. Thus, the efficient influence function (EIF) for $\psi_R$ (henceforth called $\psi$ in this subsection), under the non-parametric model, which is known to be unique, is also the efficient influence function under Assumption \ref{rosen-model}.

To obtain the EIF for $\psi = \int \E[\theta_T(X,t)]dF(t)$, we consider Equations \eqref{inf-def} and \ref{der-expand}. The second term in the Equation \eqref{der-expand} has already been computed via Lemma \ref{sec-eif}, and hence we compute $\left.\dfrac{\partial}{\partial\eta}\theta_{t'}(x,t;\eta)\right|_{\eta = 0}$. 

By the identifying Equation \eqref{mom-rosen}, we know, 
$\E_\eta\left[\psi^{t,t'}_{\theta_{t'}(x,t;\eta)}(Y)|T=t,X=x\right] = 0$, where $\E_\eta[\cdot]$ signifies that the expectation is taken under the parametric sub-model indicized by $\eta$. Recalling that $\psi^{t,t'}_\mu(y) = \max\{0,y-\mu\} + \Gamma_{t,t'}\min\{0,y-\mu\}$, we can differentiate both sides of the identifying equation with respect to $\eta$ to obtain,
\begin{align*}
    0& = \dfrac{\partial}{\partial\eta}\left[\int_{\theta_{t'}(x,t;\eta)}^\infty (y- \theta_{t'}(x,t;\eta))dF(y|x,t;\eta) + \Gamma_{t,t'}\int^{\theta_{t'}(x,t;\eta)}_{-\infty} (y- \theta_{t'}(x,t;\eta))dF(y|x,t;\eta)\right]\\
    &= \int_{\theta_{t'}(x,t;\eta)}^\infty \left[-\dfrac{\partial}{\partial\eta} \theta_{t'}(x,t;\eta)\right]dF(y|x,t;\eta) \\ &\hspace{15em} +\int_{\theta_{t'}(x,t;\eta)}^\infty (y-\theta_{t'}(x,t;\eta)) S(y|x,t;\eta)dF(y|x,t;\eta)   \\
    &\hspace{2em} +\Gamma_{t,t'}\int^{\theta_{t'}(x,t;\eta)}_{-\infty} \left[-\dfrac{\partial}{\partial\eta} \theta_{t'}(x,t;\eta)\right]dF(y|x,t;\eta) \\
    &\hspace{15em} +\Gamma_{t,t'}\int^{\theta_{t'}(x,t;\eta)}_{-\infty} (y-\theta_{t'}(x,t;\eta)) S(y|x,t;\eta)dF(y|x,t;\eta),
\end{align*}
where we have used Leibiniz's rule of differentiation of integrals. Plugging in $\eta = 0$ and compiling like terms, we obtain, 
\begin{align*}
    &\left.\dfrac{\partial}{\partial\eta}\theta_{t'}(x,t;\eta)\right|_{\eta = 0}\left(\P(Y>\theta_{t'}(x,t;0)|X=x,T=t;0)+\Gamma_{t,t'}\P(Y\le \theta_{t'}(x,t;0)|X=x,T=t;0\right) \\ &= \E\left[\psi^{t,t'}_{\theta_{t'}(x,t;0)}(Y)S(Y|x,t;0)|X=x,T=t;0\right]
\end{align*}
and thus, 
\begin{align}\label{del-theta}\left.\dfrac{\partial}{\partial\eta}\theta_{t'}(x,t;\eta)\right|_{\eta = 0} = \E\left[\left.\dfrac{\psi^{t,t'}_{\theta_{t'}(x,t;0)}(Y)S(Y|x,t;0)}{\nu_{t'}(x,t;0)}\right|X=x,T=t;0\right],
\end{align}
where $$\nu_{t'}(x,t;\eta) = \P(Y>\theta_{t'}(x,t;\eta)|X=x,T=t;\eta) + \Gamma_{t,t'}\P(Y\le \theta_{t'}(x,t;\eta)|X=x,T=t;\eta).$$
Using Equation \eqref{del-theta}, we simplify the first term on the right hand side of Equation \eqref{der-expand}.
\begin{align*}
     &\int \int \int\left.\dfrac{\partial}{\partial\eta}\theta_{t'}(x,t;\eta)\right|_{\eta = 0}dF(t'|x;0)dF(x;0)dF(t;0) \\ 
     &=  \int \int \int \E\left[\left.\dfrac{\psi^{t,t'}_{\theta_{t'}(x,t;0)}(Y)S(Y|x,t;0)}{\nu_{t'}(x,t;0)}\right|X=x,T=t;0\right]dF(t'|x;0)dF(x;0)dF(t;0) \\
     &= \int \int \int \int \dfrac{\psi^{t,t'}_{\theta_{t'}(x,t;0)}(y)}{\nu_{t'}(x,t;0)}S(y|x,t;0)dF(y|t,x;0)dF(t'|x;0)dF(x;0)dF(t;0) \\
     &= \int \int \int \left[\int\dfrac{\psi^{t,t'}_{\theta_{t'}(x,t;0)}(y)}{\nu_{t'}(x,t;0)}dF(t'|x;0)\right] S(y|x,t;0) dF(y|t,x;0)dF(x;0)dF(t;0)
\end{align*}
where the interchange of integrals is justified by Fubini's theorem, since the score function and $\theta_{t'}(X,T;0)$ admits a finite second moment, $|\psi^{t,t'}_\mu(y)|\le \Gamma_{t,t'}(Y-\mu)$, and $\nu_{t'}(x,t;0)$ is bounded below by $1$. 

Continuing with the trail of equalities, we obtain, 
\begin{align} \label{term1}
&\int \int \int \left.\dfrac{\partial}{\partial\eta}\theta_{t'}(x,t;\eta)\right|_{\eta = 0}dF(t',x;0)dF(t;0) \nonumber \\
&= \int \int \int \left[\int\dfrac{\psi^{t,t'}_{\theta_{t'}(x,t;0)}(y)}{\nu_{t'}(x,t;0)}dF(t'|x;0)\right] S(y|x,t;0) dF(y|t,x;0)dF(x;0)dF(t;0) \nonumber \\
&= \int \int \int \underbrace{\left[\int\dfrac{\psi^{t,t'}_{\theta_{t'}(x,t;0)}(y)}{\nu_{t'}(x,t;0)}dF(t'|x;0)\right]}_{=:H(y,t, x;0)} S(y|x,t;0) dF(y|t,x;0)\dfrac{dF(t;0)}{dF(t|x;0)} dF(t|x;0)dF(x;0) \nonumber \\
&= \int \int \int \left\{\dfrac{dF(t;0)}{dF(t|x;0)}H(y,t,x;0)\right\}S(y|x,t;0)dF(y|t,x;0)dF(t|x;0)dF(x;0).
\end{align}


Plugging in the expressions from the right hand side of Equation \eqref{term1} and Lemma \ref{sec-eif} onto Equation \eqref{der-expand}, we obtain,
\begin{align*}
    \left.\dfrac{d}{d\eta}\psi_\eta\right|_{\eta = 0} 
    &= \int \int \int \left\{\dfrac{dF(t;0)}{dF(t|x;0)}H(y,t,x;0)\right\}S(y|x,t;0)dF(y|t,x;0)dF(t|x;0)dF(x;0) \\
    & \hspace{2em} + \int\int \int (A_R(t,x) + B_R(t))[S(t|x;0) + S(x;0)] dF(y|t,x;0)dF(t|x;0) dF(x;0) \\ 
    &= \int \phi(w) S(w;0) dF(w)
\end{align*}
using Equation \eqref{score-split}, where $$\phi(W) = \dfrac{dF(T;0)}{dF(T|X;0)}\int \dfrac{\psi^{T,t'}_{\theta_{t'}(X,T;0)}(Y)}{\nu_{t'}(X,T;0)}dF(t'|X;0) + B_R(T) + A_R(T,X),$$
where $$A_R(T,X) = \int \theta_T(X,t)\,dF(t); \text{ and } B_R(T) = \int\int \theta_{t'}(x,T)dF(t'|x)dF(x).$$
This proves that $\phi(W)$ satisfies Equation \ref{inf-def}. Since $\E[\phi(W)] = 2\psi$ as $\E[\psi^{T,t'}_{\theta_{t'}(X,T;0)}(Y)|T,X]=0$ by the identifying equation and $A_R(T,X)$ and $B_R(T)$ both have expected value $\psi$, we thus obtain, that $\phi(W) - 2\psi$ is the efficient influence function for $\psi$ under the non-parametric model. \hfill $\blacksquare$

\subsection{Proof of Theorem \ref{mar-eif}} \label{mar-eif-proof}

By Theorem \ref{data-compatibility}, the Marginal model does not impose any restrictions on the observed data distribution. Thus the observed data tangent space under Assumption \ref{mar-model} is a superset of the tangent space under a non-parametric model, which is known to be all of $L^2$. Thus, the efficient influence function (EIF) for $\psi_M$ (henceforth called $\psi$ in this subsection), under the non-parametric model, which is known to be unique, is also the efficient influence function under Assumption \ref{mar-model}.

To obtain the EIF for $\psi = \int\E[\zeta_T(X,t)]dF(t)$, we consider Equations \eqref{inf-def} and \ref{der-expand}. The second term in the Equation \eqref{der-expand} has already been computed via Lemma \ref{sec-eif}, and hence we compute $\left.\dfrac{\partial}{\partial\eta}\zeta_{t'}(x,t;\eta)\right|_{\eta = 0}$.

First, note that, 
\begin{align*}
    & \dfrac{\partial}{\partial\eta}\E_\eta[\rho^{t,t'}_{q_{t'}(X,T;\eta)}(Y)|T=t,X=x] \\
    &= \dfrac{\partial}{\partial\eta}\left[\int \Lambda_{t,t'}^{-1}(y-q_{t'}(x,t;\eta))_+ - \Lambda_{t,t'}(y-q_{t'}(x,t;\eta))_- \,dF(y|t,x,\eta)\right]\\
    &= \dfrac{\partial}{\partial\eta}\left[\Lambda_{t,t'}^{-1}\int_{q_{t'}(x,t;\eta)}^\infty(y-q_{t'}(x,t;\eta))dF(y|t,x,\eta) + \Lambda_{t,t'}\int_{-\infty}^{q_{t'}(x,t;\eta)}(y-q_{t'}(x,t;\eta)) dF(y|t,x,\eta)\right]\\
    &= \int_{q_{t'}(x,t;\eta)}^\infty \Lambda_{t,t'}^{-1}\left(-\dfrac{\partial q_{t'}(x,t;\eta)}{\partial\eta}\right) dF(y|t,x,\eta) + \int_{-\infty}^{q_{t'}(x,t;\eta)} \Lambda_{t,t'}\left(-\dfrac{\partial q_{t'}(x,t;\eta)}{\partial\eta}\right) dF(y|t,x,\eta) \\ 
    &\hspace{2em} + \int \rho^{t,t'}_{q_{t'}(x,t;\eta)}(y) S(y|t,x,\eta) \, dF(y|t,x,\eta),
\end{align*}
using the Leibiniz's rule of differentiating integrals. Thus, plugging in $\eta = 0$ to the above, yields, 
\begin{align*}
    & \left.\dfrac{\partial}{\partial\eta}\E_\eta[\rho^{t,t'}_{q_{t'}(X,T;\eta)}(Y)|T=t,X=x]\right|_{\eta = 0} \\ &= \left(\left.-\dfrac{\partial q_{t'}(x,t;\eta)}{\partial\eta}\right|_{\eta=0}\right)\left[\Lambda_{t,t'}^{-1}\P(Y>q_{t'}(x,t)|T=t,X=x) + \Lambda_{t,t'}\P(Y\le q_{t'}(x,t)|T=t,X=x)\right] \\ &\hspace{1em} + \E[\rho^{t,t'}_{q_{t'}(x,t)}(Y)S(Y|T,X)|T=t,X=x]
\end{align*}
Since $q_{t'}(x,t)$ is a continuity point of $Y|T=t,X=x$, hence $\P(Y\le q_{t'}(x,t)|T=t,X) = (\Lambda_{t,t'}+1)^{-1}$. Thus, the term in square brackets can be simplied to 
$$\Lambda_{t,t'}^{-1}\cdot \dfrac{\Lambda_{t,t'}}{\Lambda_{t,t'}+1} + \dfrac{\Lambda_{t,t'}}{\Lambda_{t,t'}+1} = 1.$$
Hence, 
\begin{equation}
    \left.\dfrac{\partial}{\partial\eta}\E_\eta[\rho^{t,t'}_{q_{t'}(X,T;\eta)}(Y)|T=t,X=x]\right|_{\eta = 0} = \left.-\dfrac{\partial q_{t'}(x,t;\eta)}{\partial\eta}\right|_{\eta=0} + \E[\rho^{t,t'}_{q_{t'}(x,t)}(Y)S(Y|T,X)|T=t,X=x]. \label{rho-diff}
\end{equation}
Now, the first term in Equation \eqref{der-expand} requires computing the derivative of $\zeta_{t'}(x,t,\eta)$ at $\eta = 0$. That yields
\begin{align*}
    \left.\dfrac{\partial}{\partial\eta}\zeta_{t'}(x,t;\eta)\right|_{\eta=0} &= \left.\dfrac{\partial}{\partial\eta}q_{t'}(x,t;\eta)\right|_{\eta = 0} + \left.\dfrac{\partial}{\partial\eta}\E_\eta[\rho^{t,t'}_{q_{t'}(x,t;\eta)}(Y)|T=t,X=x]\right|_{\eta=0}  \\
    &= \E[\rho^{t,t'}_{q_{t'}(x,t)}(Y)S(Y|T,X)|T=t,X=x], 
\end{align*}
using Equation \eqref{rho-diff}. Thus, the first term in Equation \eqref{der-expand} can be written as 
\begin{align}
    & \int\int\int \left.\dfrac{\partial}{\partial\eta}\zeta_{t'}(x,t;\eta)\right|_{\eta=0} dF(t'|x;0)dF(x;0)dF(t;0) \nonumber \\ 
    &= \int\int\int\underbrace{\int \rho^{t,t'}_{q_{t'}(x,t)}(y) dF(t'|x)}_{=:H_1(y,x,t)} S(y|t,x)   dF(y|t,x)dF(t)dF(x) \nonumber \\
    &= \int\int\int \left[\dfrac{dF(t)}{dF(t|x)}H_1(y,x,t)\right]S(y|t,x)dF(y|t,x)dF(t|x)dF(x), \label{midway2}
\end{align}
where the interchange of integrals in the second step is justified by Fubini's theorem, and the finite conditional variance of $Y|T,X$ and the score function. Now, 
\begin{align*}\E\left[\left.\dfrac{dF(T)}{dF(T|X)}H_1(Y,X,T)\right|T=t,X=x\right] &= \dfrac{dF(t)}{dF(t|x)}\E[H(Y,x,t)|T=t,X=x] \\
&= \dfrac{dF(t)}{dF(t|x)}\int \E[\rho^{t,t'}_{q_{t'}(x,t)}(Y)|T=t,X=x]dF(t'|x) \\
&= \dfrac{dF(t)}{dF(t|x)}\int \alpha_{t'}(t,x)dF(t'|x),
\end{align*}
where $\alpha_{t'}(t,x) := \E[\rho^{t,t'}_{q_{t'}(x,t)}(Y)|T=t,X=x] = \zeta_{t'}(x,t) - q_{t'}(x,t)$ by the identifying Equation \eqref{mom-mar}. Now, since $\E[S(Y|T,X)|T=t,X=x] = 0$, we can safely plugin the above expectation in Equation \eqref{midway2} to make it mean-centered. Thus, taking $H(y,x,t) = H_1(y,x,t) - \int \alpha_{t'}(t,x)dF(t'|x)$, we have,
\begin{align}
    & \int\int\int \left.\dfrac{\partial}{\partial\eta}\zeta_{t'}(x,t;\eta)\right|_{\eta=0} dF(t'|x;0)dF(x;0)dF(t;0) \nonumber \\ &= \int\int\int \left[\dfrac{dF(t)}{dF(t|x)}H(y,x,t)\right]S(y|t,x)dF(y|t,x)dF(t|x)dF(x). \label{term1m}
\end{align}
Plugging in the expressions from the right hand side of Equation \eqref{term1m} and Lemma \ref{sec-eif} onto Equation \eqref{der-expand}, we obtain,
\begin{align*}
    \left.\dfrac{d}{d\eta}\psi_\eta\right|_{\eta = 0} 
    &= \int \int \int \left\{\dfrac{dF(t)}{dF(t|x)}H(y,x,t)\right\}S(y|x,t)dF(y|t,x)dF(t|x)dF(x) \\
    & \hspace{2em} + \int\int \int (A_M(t,x) + B_M(t))[S(t|x) + S(x)] dF(y|t,x)dF(t|x) dF(x) \\ 
    &= \int \phi(w) S(w) dF(w)
\end{align*}
using Equation \eqref{score-split}, where $$\phi(W) = \dfrac{dF(T)}{dF(T|X)}\int \left[\rho^{T,t'}_{q_{t'}(x,t)}(Y) - \alpha_{t'}(T,X)\right] dF(t'|X) + B_M(T) + A_M(T,X),$$
where $q_{t'}(x,t) = Q_{Y|X,T}((\Lambda_{T,t'}+1)^{-1})$, $$A_M(T,X) = \int \zeta_T(X,t)\,dF(t); \text{ and } B_R(T) = \int\int \zeta_{t'}(x,T)dF(t'|x)dF(x).$$
This proves that $\phi(W)$ satisfies Equation \ref{inf-def}. Since $\E[\phi(W)] = 2\psi$ as the first term is zero via debiasing through $\alpha_{t'}(t,x)$, and $A_M(T,X)$ and $B_M(T)$ both have an expected value of $2\psi$, we thus obtain $\phi(W) - 2\psi$ is the efficient influence function for $\psi$ under the non-parametric model. \hfill $\blacksquare$

\section{Proof of Asymptotic Results in Section \ref{asymp-prop}}

Let $\hat h$ as the collection of all nuisance parameters (as mentioned in Step 1), and let $H^r_{h}(t) = \E[\hat Y_r(W)|T=t, \hat h]$ and $H^m_{ h}(t) = \E[\hat Y_m(W)|T=t,\hat h]$.
\begin{lemma}[Pointwise-bias results] \label{pobias}
Suppose $\mathcal{T}$ is compact, and $Y|X=x,T=t$ has a continuous density. Then, 
    \begin{align}
        H^r_h(t) - r(t) &\lesssim_P \|\hat\theta_T(X,t) - \theta_T(X,t)\|^2_{2|\hat h} +  \|\hat\theta_T(X,t) - \theta_T(X,t)\|_{2|\hat h}\|\hat f(t|X) - f(t|X)\|_{2|\hat h} \nonumber \\ &\hspace{1em} + \|\hat\theta_T(X,t) - \theta_T(X,t)\|_{2|\hat h}\|\hat f(T|X) - f(T|X)\|_{2|\hat h} \nonumber \\ &\hspace{1em} + \|\hat\theta_T(X,t) - \theta_T(X,t)\|_{2|\hat h}\|\hat \nu_{T}(X,t) - \nu_{T}(X,t)\|_{2|\hat h} \label{pb-rosen}\\
        H^m_h(t) - m(t) &\lesssim_P \|\hat q_{T}(X,t) - q_T(X,t)\|^2_{2|\hat h} + \|\hat \zeta_{T}(X,t) - \zeta_T(X,t)\|_{2|\hat h}\|\hat f(t|X) - f(t|X)\|_{2|\hat h} \nonumber \\
        & \hspace{1em} + \|\hat \zeta_T(X,t) - \zeta_T(X,t)\|_{2|\hat h}\|\hat f(T|X) - f(T|X)\|_{2|\hat h}\label{pb-mar}.
    \end{align}
\end{lemma}
\subsection{Proof of Equation (\ref{pb-rosen}) in Lemma \ref{pobias}} \label{proof-pb-rosen}

Recall that 
$$\hat Y_r(w) = \dfrac{1}{|I_{2}|}\sum_{i\in I_{2}}  \hat\theta_{T_i}(X_i,t) + \dfrac{\dfrac{1}{|I_{2}|}\sum_{i\in I_{2}}\hat f(t|X_i)}{\hat f(t|x)}\int \dfrac{\psi^{t,t'}_{\hat\theta_{t'}(x,t)}(y)}{\hat\nu_{t'}(x,t)} \hat f(t'|x)dt'.$$

Let $\E_t$ denote the operator $\E[\cdot|T=t, \hat h]$. Thus,
\begin{align}
    H^r_h(t) - r(t) &= \E_t\left[\int\int  (\hat\theta_{t'}(x,T) - \theta_{t'}(x,T))f(t'|x)f(x) dt'dx\right]  + \E_t\left[\dfrac{\hat f(T)}{\hat f(T|X)}\int \dfrac{\psi^{T,t'}_{\hat\theta_{t'}(X,T)}(Y)}{\hat\nu_{t'}(X,T)}\hat f(t'|X)dt'\right], \label{H-expand}
\end{align}
where $\hat f(t) = \E_{\hat f}[\hat f(t|X)]$. The above equation holds since in Equation \eqref{H-expand}, $X_i$ for $i\in I_{2}$ is independent of $\{W_i\}_{i\in I_1\cup I_3}$. 
Now, 
\begin{equation} \E_t\left[\dfrac{\hat f(T)}{\hat f(T|X)}\int \dfrac{\psi^{T,t'}_{\hat\theta_{t'}(X,T)}(Y)}{\hat\nu_{t'}(X,T)}\hat f(t'|X)dt'\right] = \E_t\left[\dfrac{\hat f(t)}{\hat f(t|X)}\int \dfrac{\E_t\left[\psi^{t,t'}_{\hat\theta_{t'}(X,t)}(Y)|X\right]}{\hat\nu_{t'}(X,t)}\hat f(t'|X)dt'\right]. \label{second-term} \end{equation}
Define $$u(r) = \E_t[\psi^{t,t'}_{\theta_{t'}(X,t) + r(\hat\theta_{t'}(X,t) - \theta_{t'}(X,t))}(Y)|X],\text{ and }v(r) = \nu_{t'}(X,t) + r(\hat\nu_{t'}(X,t) - \nu_{t'}(X,t)),$$ and write $w(r) = u(r)/v(r)$, for $r\in (0,1)$. Applying Taylor theorem on $w$, we obtain,
$$w(1) = w(0) + w'(0) + w''(r^*)$$
for some $r^*\in (0,1)$. Note that, $w(0) = 0$ and $u(0) = 0$. Also,
\begin{align*} w'(r) &= \dfrac{u'(r)}{v(r)} - \dfrac{u(r)v'(r)}{v(r)^2} \\
w''(r) &= \dfrac{u''(r)}{v(r)} - 2\dfrac{u'(r)v'(r)}{v(r)^2} - \dfrac{u(r)v''(r)}{v(r)^2} + 2\dfrac{u(r)v'(r)^2}{v(r)^3}
\end{align*}
Next, for any function $g(t,X)$, define $\nu(g(t,X)) = \P(Y > g(t,X)|X,T=t) + \Gamma_{t,t'}\P(Y\le g(t,X)|X,T=t)$, so that $\nu_{t'}(X,t) = \nu(\theta_{t'}(X,t))$. Under such notation, one obtains,
\begin{align*}
    u'(r) &= -(\hat\theta_{t'}(X,t) - \theta_{t'}(X,t))\nu(\theta_{t'}(X,t) + r(\hat\theta_{t'}(X,t) - \theta_{t'}(X,t))) \\
    u''(r) &= -(\hat\theta_{t'}(X,t) - \theta_{t'}(X,t))^2(\Gamma_{t,t'} - 1)f_{Y|X,T=t}(u(r))
\end{align*}
Also, $v'(r) = (\hat\nu_{t'}(X,t) - \nu_{t'}(X,t))$, and $v''(r) = 0$. Thus,
$$w'(0) = \dfrac{u'(0)}{v(0)} - \dfrac{u(0)v'(0)}{v(0)^2} = - (\hat\theta_{t'}(X,t) - \theta_{t'}(X,t)) $$ noting that $u'(0) = -(\hat\theta_{t'}(X,t) - \theta_{t'}(X,t))v(0)$ and $u(0) = 0$. Also,
$v''(r) = 0$ and $u'(r) = -v(r)(\hat\theta_{t'}(X,t) - \theta_{t'}(X,t))$ leads to 
\begin{align} w''(r) &= \dfrac{u''(r)}{v(r)} + 2(\hat\theta_{t'}(X,t) - \theta_{t'}(X,t))\dfrac{v'(r)}{v(r)} + 2\dfrac{u(r)v'(r)^2}{v(r)^3} \nonumber \\
&= -(\hat\theta_{t'}(X,t) - \theta_{t'}(X,t))^2\dfrac{(\Gamma_{t,t'} - 1)f_{Y|X,T=t}(\theta_{t'}(X,t) + r(\hat\theta_{t'}(X,t) - \theta_{t'}(X,t)))}{v(r)} \nonumber \\ & \hspace{2em}+ 2(\hat\theta_{t'}(X,t) - \theta_{t'}(X,t))(\hat\nu_{t'}(X,t) - \nu_{t'}(X,t))\dfrac{1}{v(r)} + 2(\hat\nu_{t'}(X,t) - \nu_{t'}(X,t))^2\dfrac{u(r)}{v(r)^3} \label{doublederivative}
\end{align}
Thus, from Equation \eqref{second-term}, we obtain, 
\begin{align}
    \E_t\left[\left.\int \dfrac{\psi^{T,t'}_{\hat\theta_{t'}(X,T)}(Y)}{\hat\nu_{t'}(X,T)}\hat f(t'|X)dt'\right|X\right]&= \int w(1)\hat f(t'|X) dt' = \int w'(0) \hat f(t'|X)dt' + \int w''(r^*) \hat f(t'|X)dt'\nonumber  \\
    &= -\int (\hat\theta_{t'}(X,t) - \theta_{t'}(X,t))\hat f(t'|X)dt' + \int w''(r^*)\hat f(t'|X)dt' \label{-w-double}
\end{align}
Now, 
\begin{align*} &-\E_t\left[\dfrac{\hat f(t)}{\hat f(t|X)}\int (\hat\theta_{t'}(X,t) - \theta_{t'}(X,t))\hat f(t'|X)dt'\right] \\ &= -\int\int \dfrac{\hat f(t)}{\hat f(t|x)}(\hat\theta_{t'}(x,t) - \theta_{t'}(x,t))\hat f(t'|x) dt' f(x|t)\,dx \\
&= -\int\int (\hat\theta_{t'}(x,t) - \theta_{t'}(x,t))\left(\dfrac{\hat f(t)}{f(t)}\cdot \dfrac{\ f(t|x)}{\hat f(t|x)}\cdot \dfrac{\hat f(t'|x)}{f(t'|x)}\right)f(t'|x) f(x)dt'dx 
\end{align*}
Hence, from Equations \eqref{H-expand} and \eqref{-w-double}, we have, 
\begin{align}
    H^r_h(t) - r(t) &= \int\int (\hat\theta_{t'}(x,t) - \theta_{t'}(x,t))\left(1-\dfrac{\hat f(t)}{f(t)}\cdot \dfrac{f(t|x)}{\hat f(t|x)}\dfrac{\hat f(t'|x)}{f(t'|x)}\right) f(t'|x) f(x)dt'dx \nonumber \\ &\hspace{2em} + \E_t\left[\dfrac{\hat f(T)}{\hat f(T|X)}\int w''(r^*) \hat f(t'|X)\,dt'\right] \label{midway}
\end{align}
The first term on the right hand side of Equation \eqref{midway} can be simplified as 
\begin{align*}
    &\int\int (\hat\theta_{t'}(x,t) - \theta_{t'}(x,t))\left(1-\dfrac{\hat f(t)}{f(t)}\cdot \dfrac{f(t|x)}{\hat f(t|x)}\dfrac{\hat f(t'|x)}{f(t'|x)}\right) f(t'|x)f(x)dt'dx \\ &= \E_{\hat h}\left[(\hat\theta_T(X,t) - \theta_T(X,t))\left(1- \dfrac{\hat f(t)}{f(t)}\dfrac{f(t|X)}{\hat f(t|X)}\dfrac{\hat f(T|X)}{f(T|X)}\right)\right].
\end{align*}
Continuing with the trail of equalities, and noting that 
$$1- \dfrac{\hat f(t)}{f(t)}\dfrac{f(t|X)}{\hat f(t|X)}\dfrac{\hat f(T|X)}{f(T|X)} = \dfrac{f(T|X) - \hat f(T|X)}{f(T|X)} + \dfrac{\hat f(T|X)}{f(T|X)}\left[\dfrac{1}{f(t)}(f(t) - \hat f(t)) + \dfrac{\hat f(t)}{f(t)\hat f(t|X)}(\hat f(t|X) - f(t|X))\right],$$ and $$|\hat f(t) - f(t)| = \left|\E_{\hat h}[\hat f(t|X) - f(t|X)]\right| \le \|\hat f(t|X) - f(t|X)\|_{2|\hat h},$$ we obtain that the first term on the right hand side of Equation \eqref{midway} is 
\begin{align}
    &\lesssim_P  (f(t)-\hat f(t))\E_{\hat h}\left[\hat\theta_T(X,t) - \theta_T(X,t)\right] + \E_{\hat h}\left[\left(\hat\theta_T(X,t) - \theta_T(X,t)\right)\left(\hat f(t|X) - f(t|X)\right)\right] \nonumber \\ &\hspace{2em} + \E_{\hat h}[(\hat\theta_T(X,t) - \theta_T(X,t))(\hat f(T|X) - f(T|X))] \nonumber \\
    &\lesssim_P \|\hat\theta_T(X,t) - \theta_T(X,t)\|_{2|\hat h}\|\hat f(t|X) - f(t|X)\|_{2|\hat h} + \|\hat\theta_T(X,t) - \theta_T(X,t)\|_{2|\hat h}\|\hat f(T|X) - f(T|X)\|_{2|\hat h}. \label{first-component}
\end{align}
Next, we need to control the second term on the right hand side of Equation \eqref{midway}. Note that, for any $r$, $v(r)\in [1,\Gamma_{T,t'}]$;  $\Gamma_{T,t'}$ is bounded above since $T$ is assumed to have a compact support, and the density of $Y|X,T=t$ is assumed to be bounded. Also, by Lipschitz continuity of $s\mapsto \psi^{t,t'}_s(Y)$, we have, $|u(r)|\le r\Gamma_{t,t'}(\hat\theta_{t'}(X,t) - \theta_{t'}(X,t))$. Thus, from Equation \eqref{doublederivative}, we have, 
\begin{align} 
|w''(r)| &\lesssim (\hat\theta_{t'}(X,t) - \theta_{t'}(X,t))^2 + |(\hat\theta_{t'}(X,t) - \theta_{t'}(X,t))(\hat\nu_{t'}(X,t) - \nu_{t'}(X,t))|  \nonumber \\ &\hspace{2em}  + (\hat\nu_{t'}(X,t) - \nu_{t'}(X,t))^2|\hat\theta_{t'}(X,t) - \theta_{t'}(X,t)|. \label{w-bound}
\end{align}
Now, for any function $a(t',t,x)$, 
\begin{align*}
    \E_t\left[\dfrac{\hat f(T)}{\hat f(T|X)}\int a(t',T,X) \hat f(t'|X)\,dt'\right] &= \int \int \dfrac{\hat f(t)}{\hat f(t|x)} a(t',t,x) \hat f(t'|x) f(x|t)\,dt'\,dx \\
    &= \int\int \left(\dfrac{f(t|x)}{\hat f(t|x)}\cdot \dfrac{\hat f(t)}{f(t)}\cdot \dfrac{\hat f(t'|x)}{f(t'|x)}\right) a(t',t,x) f(t'|x)f(x)\,dt'\,dx \\
    &\lesssim \E[a(T,t,X)]
\end{align*}
Thus, from Equations \eqref{-w-double} and \eqref{w-bound}, we obtain that the second term on the right hand side of Equation \eqref{midway} is 
\begin{align}
    & \lesssim_P \E_{\hat h}(\hat\theta_{T}(X,t) - \theta_{T}(X,t))^2 + \E_{\hat h}(\hat\theta_{T}(X,t) - \theta_{T}(X,t))(\hat\nu_{T}(X,t) - \nu_{T}(X,t)) \nonumber \\
    &\le \|\hat\theta_{T}(X,t) - \theta_{T}(X,t))^2\|^2_{2|\hat h} + \|\hat\theta_{T}(X,t) - \theta_{T}(X,t)\|_{2|\hat h}\|\hat\nu_{T}(X,t) - \nu_{T}(X,t)\|_{2|\hat h} \label{second-component}
\end{align}
The proof is complete by noting that the right hand side of Equation \eqref{midway} is the sum of the components of Equations \eqref{first-component} and \eqref{second-component}.

\subsection{Proof of Equation (\ref{pb-mar}) in Lemma \ref{pobias}} 

Define $\alpha_{t'}(x,t) = \zeta_{t'}(x,t) - q_{t'}(x,t)$, and recall that 
$$\hat Y_m(w) = \dfrac{1}{|I_2|}\sum_{j\in I_2} \hat\zeta_{T_j}(X_j,t) + \dfrac{\dfrac{1}{|I_2|}\sum_{j\in I_2}\hat f(t|X_j)}{\hat f(t|x)}\int [\rho^{t,t'}_{\hat q_{t'}(x,t)}(y) - \hat\alpha_{t'}(x,t)]\hat f(t'|x)dt'.$$
As before, denote $\E_t$ as the operator $\E[\cdot|T=t,\hat h]$. Thus,
\begin{align} H_h^m(t) - m(t) &=\E_t\left[\int\int (\hat\zeta_{t'}(x,T) - \zeta_{t'}(x,T))f(t'|x)f(x) dt'dx\right] \nonumber \\ &\hspace{1em} + \E_t\left[\dfrac{\hat f(t)}{\hat f(t|X)}\int \left\{\E_t[\rho^{t,t'}_{\hat q_{t'}(X,t)}(Y)|X] - \hat\alpha_{t'}(X,t)\right\}\hat f(t'|X)dt'\right] \label{H-expand-2}
\end{align}
where $\hat f(t) = \E_{\hat f}[\hat f(t|X)]$, and Equation \eqref{H-expand-2} is justified as $X_i$ for $i\in I_2$ is independent of $\{W_i\}_{i\in I_1\cup I_3}$. 

Now, define $$u(r) = \E_t[\rho^{t,t'}_{q_{t'}(X,t) + r(\hat q_{t'}(X,t)-  q_{t'}(X,t))}(Y)|X],\text{ and }v(r) = \alpha_{t'}(X,t) + r(\hat\alpha_{t'}(X,t) - \alpha_{t'}(X,t)),$$
and write $w(r) = u(r) - v(r)$, for $r\in (0,1)$. Applying Taylor theorem on $w$ we obtain,
$$w(1) = w(0) + w'(0) + w''(r^*)$$ for some $r^*\in (0,1)$. Note that, $w(0) = 0$ as $u(0) = v(0)$. Moreover, 
\begin{align*}
    u'(r) &= -\E_t[(\hat q_{t'}(X,t) - q_{t'}(X,t))(\Lambda_{t,t'}^{-1}\P(Y>q_{t'}(X,t) + r(\hat q_{t'}(X,t) - q_{t'}(X,t))|T=t,X) \\ & \hspace{1em} + \Lambda_{t,t'}\P(Y\le q_{t'}(X,t) + r(\hat q_{t'}(X,t) - q_{t'}(X,t))|T=t,X))]\\
    v'(r)&= \hat\alpha_{t'}(X,t) -\alpha_{t'}(X,t).
\end{align*}
Thus, 
\begin{align*}
    w'(0) &= u'(0) - v'(0) \\
    &= -[\hat q_{t'}(X,t) - q_{t'}(X,t)] - (\hat\alpha_{t'}(X,t)-\alpha_{t'}(X,t)),
\end{align*}
using the fact that $\P(Y\le q_{t'}(X,t)|X,T=t) = (\Lambda_{t,t'}+1)^{-1}.$
Noting that $q_{t'}(X,t) + \alpha_{t'}(X,t) = \zeta_{t'}(X,t)$, we thus have, 
$w'(0) = -\int (\hat \zeta_{t'}(X,t) - \zeta_{t'}(X,t)) \hat f(t'|X)\,dt'$. Moreover, since the density of $Y|T=t,X=x$ is bounded, hence 
\begin{align} 
|w''(r^*)| &= |u''(r^*)| \lesssim (\Lambda_{t,t'}-\Lambda_{t,t'}^{-1})\E_t[(\hat q_{t'}(X,t) - q_{t'}(X,t))^2|X] \nonumber \\
& \lesssim (\hat q_{t'}(X,t) - q_{t'}(X,t))^2 \label{w-bound-2}
\end{align}
where the last line follows from the compact support of $T$. Thus, the second term of Equation \eqref{H-expand-2} can be written as 
\begin{align} -\E_t\left[\dfrac{\hat f(t)}{f(t|X)}\int w(1) \hat f(t'|X)dt'\right] &= -\E_t\left[\dfrac{\hat f(t)}{f(t|x)}\int (\hat\zeta_{t'}(X,t) -\zeta_{t'}(X,t))\hat f(t'|X)dt'\right] \nonumber \\ & \hspace{1em}+ \E_t\left[\dfrac{\hat f(t)}{f(t|X)}\int w''(r^*)\hat f(t'|X)dt'\right] \label{midway-mar}.
\end{align}
Since Equation \eqref{midway-mar} leads us to an equation remarkably similar to Equation \eqref{midway}, the proof follows by following analogous following steps as in Subsection \ref{proof-pb-rosen}, and noting the bounds in Equation \eqref{w-bound-2}.


\subsection{Proof of Result in Theorem \ref{l2-cons}: \texorpdfstring{$L_2$}{L2}-consistency} \label{l2-cons-proof}

Let $H_h^a(t) = \E[\hat Y_a(W)|T=t,\hat h]$, where $a=\{r,m\}$.

\textbf{Step 1: Components of Counterfactual Regression}
If $\xi_k^2\log k/n\to 0$, then from  \cite{belloni2015some} we obtain, 
\begin{equation} \|H_h^a - \hat a_J\|_{2|\hat h} := \E[(H_h^a(T') - \hat a_J(T'))^2|\hat h]^{\frac 12} = O_p\left(\sqrt{\dfrac{\bar\sigma^2_aJ}{|I_1|}} + c_J\right) \label{belloni-cons} \end{equation}
where $\bar\sigma^2$ is an upper bound on $\V[\hat Y_a(W)|T',\hat h]$, and 
$$c_J = \inf_{b\in \R^J}\E[(H_h^a(T') - b^T\bar\phi_J(T'))^2|\hat h]^{\frac 12} = \inf_{b\in \R^J} \|H_h^a - b^T\bar\phi_J\|_{2|\hat h}$$

Now, for any $b\in \R^J$, we have, 
$$\|H_h^a - b^T\bar\phi_J\|_{2|\hat h} \le \|H_h^a - a\|_{2|\hat h} + \|a - b^T\bar \phi_J\|_{2|\hat h},$$
and noting that the second term on the right hand side does not depend on $\hat h$, one obtains, 
\begin{align} c_J &\le \|H_h^a - a\|_{2|\hat h} + \inf_{b\in \R^J} \E[(a(T') - b^T\bar\phi_J(T'))^2]^{\frac 12} \nonumber  \\
&\le \|H_h^a - a\|_{2|\hat h} + \kappa E_J^\Psi(a) \label{c_J}
\end{align}
where the last inequality follows since $T$ is assumed to have a density on its support with respect to the Lebesgue measure $\lambda$, which is bounded above by a constant, say $\kappa$.

Thus, from Equation \eqref{belloni-cons}, we obtain
\begin{align*}
    \|\hat a_J - a\|_{2|\hat h} &\le \|\hat a_J - H^a_h\|_{2|\hat h} + \|H_h^a - a\|_{2|\hat h} 
    = O_p\left(\sqrt{\dfrac{{\bar\sigma^2_aJ}}{|I_1|}} + c_J\right) + \|H_h^a - a\|_{2|\hat h} \\
    & = O_p\left(\sqrt{\dfrac{\bar\sigma^2_a J}{|I_1|}} + E_J^\Psi(a) + \|H_h^a-a\|_{2|\hat h}\right),
\end{align*}
where the last equality follows from Equation \eqref{c_J}. This establishes the first two terms in Equations \eqref{rosen-l2} and \eqref{mar-l2}. The third term quantifies the contribution from estimating the nuisance functions, which we describe in Step 2, in conjunction with Lemma \ref{pobias}.


\textbf{Step 2: Bias from estimation of nuisance functions}

$\|H_h^a - a\|^2_{2|\hat h} = \int (H_h^a(t') - a(t'))^2_{2|\hat h} dF_T(t')$

Now, if $|H_h^a(t') - a(t')| = \sum_{k=1}^K d_k(t')$, then by Cauchy-Schwarz inequality, we have, 
$$\|H_h^a - a\|^2_{2|\hat h} = \int \left(\sum_{k=1}^K d_k(t')\right)^2dF_T(t') \le K\sum_{k=1}^K\int d_k(t')^2dF_T(t).$$
Thus, we need to control the second moments of each individual terms of $H_h^a - a$. 

Now, by Cauchy-Schwarz inequality,
$$\int (\E_{\hat h}[\hat\theta_T(X,t') - \theta_T(X,t')^2])^2 dF_T(t')\le \int  \E_{\hat h}[(\hat\theta_T(X,t') - \theta_T(X,t'))^4]dF_T(t') = \|\hat\theta_T(X,T') - \theta_T(X,T')\|_{4|\hat h}^4.$$
Analogous expressions hold to establish $\|\hat q_T(X,T') - q_T(X,T')\|^4_{4|\hat h}$.

Next, for any function $h_1(T,X,t')$ and $h_2(T,X,t')$, 
\begin{align*}
    & \int (\E[(\hat h_1(T,X,t') - h_1(T,X,t'))(\hat h_2(T,X,t') - h_2(T,X,t'))])^2 dF_T(t') \\ 
    &\le \int \E[(\hat h_1(T,X,t') - h_1(T,X,t'))^2]\E[(\hat h_2(T,X,t') - h_2(T,X,t'))^2] dF_T(t')\\
    &\le \|\hat h_1(T,X,T') - h_1(T,X,T')\|^2_{4|\hat h}\|\hat h_2(T,X,T') - h_2(T,X,T')\|^2_{4|\hat h}
\end{align*}
by repeatedly applying Cauchy-Schwarz inequality. Taking $h_1$ to be $\theta$ or $\nu$ or $\zeta$ and $h_2$ to be $f$ produces desired results.

Next, note that,
\begin{align*}
    \|\hat f(T'|X) - f(T'|X)\|^4_{4|\hat h} &= \int (\hat f(t'|x) - f(t'|x))^4 dF_T(t')dF_X(x) \\ &=  \int (\hat f(t'|x) - f(t'|x))^4 \dfrac{f_T(t')}{f_{T|X}(t'|x)}dF_{T|X}(t')dF_X(x) \\
    &\le \dfrac{1-p_{\m}}{p_\m}\int (\hat f(t'|x) - f(t'|x))^4 dF_{T|X}(t')dF_X(x) \lesssim \|\hat f(T|X) - f(T|X)\|^4_{4|\hat h}
\end{align*}

Collecting the respective components, and noting that $\sqrt{\sum_{k=1}^K d_K^2}\le \sum_{k=1}^K |d_K|$, the proof is complete. \hfill $\blacksquare$





\subsection{Proof of Theorem \ref{asymp-norm}: Pointwise Asymptotic Normality} \label{asymp-norm-proof}

Recall that, 
$$\hat a_J(t) = \bar\phi_J(t)^T \hat Q_{|I_1|}^{-1}\cdot \dfrac{1}{|I_1|}\sum_{i\in I_1} \bar\phi_J(T_i)\hat Y_a(W_i),$$ 
where $\hat Q_{|I_1|} = \dfrac{1}{|I_1|}\sum_{i\in I_1}\bar\phi_J(T_i)\bar\phi_J(T_i)^T$, and $a\in \{r,m\}$. Consider the estimator which one could have obtained if the pseudo-outcome $Y_a(w)$ were known ---
$$a_J(t) := \bar\phi_J(t)^T\hat Q_{|I_1|}^{-1}\cdot \dfrac{1}{|I_1|}\sum_{i\in I_1}\bar\phi_J(T_i)Y_a(W_i).$$
Then, one can write 
\begin{equation}
    \sqrt{|I_1|}(\hat a_J(t) - a(t)) = \sqrt{|I_1|}(\hat a_J(t) - a_J(t)) + \sqrt{|I_1|}(a_J(t) - a(t)). \label{split-equation}
\end{equation}
Now,
\begin{align}
    \hat a_J(t) - a_J(t) &= \bar\phi_J(t)^T\hat Q^{-1}_{|I_1|}\cdot \dfrac{1}{|I_1|}\sum_{i\in I_1}\bar\phi_J(T_i)(\hat Y_a(W_i) - Y_a(W_i)) \nonumber \\ 
    &= \bar\phi_J(t)^T\hat Q^{-1}_{|I_1|}\cdot \left(\dfrac{1}{|I_1|}\sum_{i\in I_1}\bar\phi_J(T_i)(\hat Y_a(W_i) - Y_a(W_i)) - \E_{\hat h}(\bar\phi_J(T)(\hat Y_a(W) - Y_a(W)))\right) \tag{\text{EP}} \label{ep} \\ &\hspace{1em} + \bar\phi_J(t)^T\hat Q^{-1}_{|I_1|}\E_{\hat h}(\bar\phi_J(T)(\hat Y_a(W) - Y_a(W))) \tag{\text{Bias}}. \label{bp}
\end{align}    

We shall control the empirical process term \eqref{ep} and the bias term \eqref{bp} separately. This will allow us to demonstrate that the asymptotic normality is driven by the first term in Equation \eqref{split-equation}, henceforth referred to as the driving term. 

\textbf{Step 1: Empirical Process Term}

To control the empirical process term, note that 
\begin{align}
    &\left|\bar\phi_J(t)^T\hat Q_{|I_1|}^{-1}\cdot \left(\dfrac{1}{|I_1|}\sum_{i\in I_1}\bar\phi_J(T_i)(\hat Y_a(W_i) - Y_a(W_i)) - \E_{\hat h}(\bar\phi_J(T)(\hat Y_a(W) - Y_a(W)))\right) \right| \nonumber \\
    &\le 
    \|\bar\phi_J(t)\|\left\|\hat Q_{|I_1|}^{-1}\cdot \left(\dfrac{1}{|I_1|}\sum_{i\in I_1}\bar\phi_J(T_i)(\hat Y_a(W_i) - Y_a(W_i)) - \E_{\hat h}(\bar\phi_J(T)(\hat Y_a(W) - Y_a(W)))\right)\right\|. \label{diff-pseudo}
\end{align}
Now, since $\xi_J^2\log J/n\to 0$ by Assumption \ref{tecn}(iii), one can apply matrix Law of Large numbers (Lemma 6.2 of \cite{belloni2015some}) to obtain that $\|\hat Q - Q\|\overset{p}{\to}0$. Since all eigen values of $Q = \E[\bar\phi_J(T)\bar\phi_J(T)^T]$ are bounded away from 0, by following the argument of the proof of Theorem 4.1 of \cite{belloni2015some}, one can obtain from Equation \eqref{diff-pseudo} that
\begin{align}
    &\left\|\hat Q_{|I_1|}^{-1}\cdot \left(\dfrac{1}{|I_1|}\sum_{i\in I_1}\bar\phi_J(T_i)(\hat Y_a(W_i) - Y_a(W_i)) - \E_{\hat h}(\bar\phi_J(T)(\hat Y_a(W) - Y_a(W)))\right)\right\| \nonumber \\ &\lesssim_P \left\|\dfrac{1}{|I_1|}\sum_{i\in I_1}\bar\phi_J(T_i)(\hat Y_a(W_i) - Y_a(W_i)) - \E_{\hat h}(\bar\phi_J(T)(\hat Y_a(W) - Y_a(W)))\right\|. \label{qbound}
\end{align}

Since the right hand side of Equation \eqref{qbound}, conditional on $\hat h$, is a mean of iid terms, we can thus apply Lemma 2 of the supplementary materials of \cite{kennedysharp2020} to obtain
\begin{equation*}
    \left\|\hat Q_{|I_1|}^{-1}\cdot \left(\dfrac{1}{|I_1|}\sum_{i\in I_1}\bar\phi_J(T_i)(\hat Y_a(W_i) - Y_a(W_i)) - \E_{\hat h}(\bar\phi_J(T)(\hat Y_a(W) - Y_a(W)))\right)\right\| \lesssim_P \dfrac{1 }{\sqrt{|I_1|}}\|\bar\phi_J(T)(\hat Y_a - Y_a)\|_{2|\hat h}. 
\end{equation*}
Now, $\|\bar\phi_J(T)(\hat Y_a-Y_a)\|_{2|\hat h}\le \sup_{t\in\T}\|\bar \phi_J(t)\|\|\hat Y_a - Y_a\|_{2|\hat h}\le \xi_J\|\hat Y_a-Y_a\|_{2|\hat h}$. Moreover, $\|\bar \phi(t)\|\le \lambda_{\max}(\Omega)^{\frac 12}\|s(t)\|$. Thus, 
 we obtain that the empirical process term,
\begin{equation}
    \ref{ep} = O_p\left(\dfrac{\|s(t)\|\xi_J}{\sqrt{|I_1|}}\|\hat Y_a - Y_a\|_{2|\hat h}\right). \label{ep-final}
\end{equation}
Since $\|\hat h - h\|_{2|\hat h} = o_p(1/\xi_J)$, a quick calculation for $a\in\{r,m\}$ reveals $\xi_J\|\hat Y_a - Y_a\|_{2|\hat h}\lesssim_P \xi_J\|\hat h - h\|_{2|\hat h} = o_p(1)$, and thus from Equation \eqref{ep-final}, $\ref{ep} = o_p(\|s(t)\|/\sqrt{|I_1|})$.


\textbf{Step 2: Bias Term}

To control the bias term \eqref{bp}, we note that,
\begin{align*}\E_{\hat h}[\bar\phi_J(T)(\hat Y_a(W) - Y_a(W))] &= \E_{\hat h}[\bar\phi_J(T)\E_T[\hat Y_a(W) - Y_a(W)]] = \E_{\hat h}[\bar\phi_J(T)(H_h^a(T) - a(T)))]. 
\end{align*}
Also, using the same matrix LLN argument as while controlling the empirical process term, we can bound the bias term by the following.
\begin{align}
    \ref{bp} &\le \|\bar\phi_J(t)\|\|\hat Q^{-1}_{|I_1|}\E_{\hat h}(\bar\phi_J(T)(\hat Y_a(W) - Y_a(W)))\| \nonumber \\ &\lesssim_P \|\bar\phi_J(t)\|  \|\E_{\hat h}(\bar\phi_J(T)(\hat Y_a(W) - Y_a(W))\| \le \|\bar\phi_J(t)\|\|\E_{\hat h}[\bar\phi_J(T) (H_h^a(T) - a(T))]\|\nonumber \\ &\lesssim_P \|s(t)\|  \|\phi_J(T)\|_2 \|H_h(T) - m(T)\|_{2|\hat h} \label{bp-final}
\end{align}
Now,
\begin{equation} \label{e-bound} \|\bar \phi_J(T)\|_2^2 =\E[\bar\phi_J(T)^T\bar \phi_J(T)] =\text{tr}(Q) \le J l_{\max} \implies \|\bar\phi_J(T)\|_2 \lesssim \sqrt J.
\end{equation}
Following steps analogous to Step 2 of Appendix \ref{l2-cons-proof}, and using Equation \eqref{e-bound} we thus obtain $\ref{bp} = o_p(\|s(t)\|/\sqrt{|I_1|})$.

\textbf{Step 3: Driving term}


For $\sqrt{|I|_1}(a_J(t) - a(t))$ we can apply the asymptotic normality result  by \cite{belloni2015some}, 
\begin{equation}
    \dfrac{\sqrt{|I_1|}}{\|s(t)\|}(m_J(t) - m(t)) \overset{d}{\to} \mathcal{N}(0,1), \label{main-normal}
\end{equation}
if the following conditions hold:
\begin{enumerate}[label = (\roman*)]
    \item (Invertibility and finiteness condition) The eigen-values of $Q = \E[\bar\phi_J(T)\bar\phi_J(T)]^T$ are bounded above and away from 0.
    \item (Lindeberg condition) With $\varepsilon_a = Y_a(W) - a(T)$ is such that $$\sup_{t\in\mathcal{T}}\E[\varepsilon_a^2\mathbbm{1}(|\varepsilon_a|>M)|T=t]\to 0\text{ as }M\to\infty.$$
    \item (Non-degeneracy condition) $\exists C>0$ such that $\inf_{t\in\mathcal{T}} \E[\varepsilon_a^2|T=t]\ge C.$
    \item (Basis condition) $\sqrt{\dfrac{\xi_J^2\log J}{n}}(1+\sqrt{J}l_J)\to 0.$ 
    \item (Under-smoothing) Let $e(t) = a(t) - \bar\phi_J(t)^T\beta^a_J$. 
    Then, $\sqrt{n}e(t) = o(\|s(t)\|).$
\end{enumerate}

(i) holds by Assumption \ref{tecn}(i). 

To show (ii), note that 
$$\E[|\varepsilon_a|^{2+\delta}|T=t] \ge \E[|\varepsilon_a|^{2+\delta}\mathbbm{1}(|\varepsilon_a|>M)|T=t] \ge M^\delta \E[\varepsilon_a^2\mathbbm{1}(|\varepsilon_a|>M)|T=t],$$ and hence for (ii) to hold it is enough that $\sup_t\|\varepsilon_a\|_{2+\delta|t}$ is finite. Now, by Minkowski's inequality, we obtain,
$$\|\varepsilon_a\|_{2+\delta|T=t} = \|Y_a(W) - a(T)\|_{2+\delta|T=t}\le \|Y_a(W)\|_{2+\delta|T=t} + a(t).$$

Now, if $a = r$, 
\begin{align*} \|Y_r(W)\|_{2+\delta|T=t} \le \|\theta_T(X,t)\|_{2+\delta} + \left\|\dfrac{f_T(t)}{f_{T|X}(t|X)}\int \dfrac{\psi^{t,t'}_{\theta_{t'}(X,t)}(Y)}{\nu_{t'}(X,t)}f_{T|X}(t'|X)dt'\right\|_{2+\delta|T=t}
\end{align*}
Next note that $f_{T|X}$ and hence $f_T$ are bounded above and below uniformly in $\mathcal{T}$, and $\nu_{t'}(X,t)\ge 1$. Moreover, $\psi^{t,t'}_\theta(Y) \le \Gamma_{t,t'}(y-\theta)$, we thus have, the second term to be 
$$\lesssim_P \|Y - \theta_{t'}(X,t)\|_{2+\delta|T=t} \le \|Y\|_{2+\delta|T=t} + \|\theta_{t'}(X,t)\|_{2+\delta|T=t}.$$
(ii) thus follows by taking supremum over $t$, and considering Assumption \ref{tecn-2} (ii). One can show $\sup_{t\in \T}\|Y_m(W)\|_{2+\delta|T=t}$ analogously under Assumption \ref{tecn-2} (ii). 

To show (iii), first let's consider when $a=r$. Then,
$$Y_r(W) - r(T) = \dfrac{1}{|I_2|}\sum_{j\in I_2} (\theta_{T_j}(X_j,T) - \E_T[\theta_{T_j}(X_j,T)]) + \E_T[\theta_{T_j}(X_j,T)] + \dfrac{f(T)}{f(T|X)}\int \dfrac{\psi^{T,t'}_{\theta_{t'}(X,T)}(Y)}{\nu_{t'}(X,T)} f(t'|X)dt'.$$
The first and third terms are mean-centered conditional on $T$, and all three terms are independent conditional on $T$. Hence, 
$$\V[Y_r(W) - r(T)|T=t]\ge \V(\theta_{T}(X,t))>0$$ 
by Assumption \ref{tecn-2} (iii). The case for $a=m$ is similar.

(iv) holds by Assumptions \ref{tecn}(iii) and \ref{tecn-2}(iv). (v) is a restatement of the under-smoothing condition in Theorem \ref{asymp-norm}.


Thus, Equation \eqref{main-normal} holds, and hence the proof is complete by reconciling Equations \eqref{split-equation}, \eqref{ep-final} and \eqref{bp-final}. \hfill $\blacksquare$

\subsection{Variance Estimation} \label{var-est-proof}

\textbf{Regularity conditions required for Theorem \ref{var-est}}:

\begin{itemize}
    \item[(i)] $\xi_J^{1+\frac 4\delta}\log J\lesssim n$ 
    \item[(ii)] $\log\xi_J^L \lesssim_P \log J$, where $\xi_J^L = \sup_{t,t'\in\T,t\ne t'} \dfrac{\|\frac{\bar\phi_J(t)}{\bar\phi_J(t)} - \frac{\bar\phi_J(t')}{\bar\phi_J(t')} \|}{\|t-t'\|}$.
    \item[(iii)] $\log\xi_J\lesssim \log J$.
    \item[(iv)] $\sqrt{\dfrac{\xi_J^2\log J}{|I_1|}}(|I_1|^{1/(2+\delta)}\sqrt{\log J} + \sqrt{J}l_J)+ \sqrt{\log J}l_J \lesssim \log J$.
    \item[(v)]  $\lambda_h:= \max_{i\in |I_1|}|\hat h(W_i)-h(W_i)|$  is such that $\lambda_h(|I_1|^{\frac{1}{2+\delta}}+ l_J) =o_p(1).$
    \item[(vi)] $\|\hat h - h\|_{\infty|\hat h} = o_p(1).$
\end{itemize}

\textit{Proof of Theorem \ref{var-est}:}

Let $\hat Z = \hat\varepsilon_a \bar\phi_J(T)$, and $Z = \varepsilon_a \bar\phi_J(T)$, where $\varepsilon_a = Y_a(W) - a(T)$, and $\hat\varepsilon_a = \hat Y_a(W) - \hat a_J(T)$. Furthermore, define $\tilde\varepsilon_a = Y_a(W) - a_J(T)$, and $\tilde Z = \tilde\varepsilon_a \bar\phi_J(T)$.

We shall use the notation $\E_n$ to denote the empirical average, as $\E_nZ = \frac{1}{|I_1|}\sum_{i\in |I_1|} Z_i$, and $\E Z = \E[Z|\hat h]$. Then, we have,
$$\E_n\hat Z\hat Z^T - \E ZZ^T = \E_n (\hat Z\hat Z^T -  \tilde Z\tilde Z^T) + \E_n \tilde Z\tilde Z^T - \E ZZ^T.$$
Now, by \cite{belloni2015some} Theorem 4.6, we already have 
$$\|\E_n \tilde Z\tilde Z^T - \E ZZ^T\|\lesssim_P (|I_1|^{1/(2+\delta)} + l_J)\sqrt{\dfrac{\xi_J^2\log \xi_J}{|I_1|}}.$$
Thus, we need to control the term $\E_n (\hat Z\hat Z^T -\tilde Z\tilde Z^T).$ Note that, 
$$(\hat Z\hat Z^T -\tilde Z\tilde Z^T) = \phi_J(T)\phi_J(T)^T [(\hat Y_a - \hat a_J(T))^2 - (Y_a - a_J(T))^2].$$
Also,
\begin{align*}
    (\hat Y_a - \hat a_J(T))^2 - (Y_a - a_J(T))^2 &= [(\hat Y_a-\hat a_J(T)) - (Y_a - a_J(T))]^2 \\ &\hspace{1em} + 2(Y_a-a_J(T))[(\hat Y_a-\hat a_J(T)) - (Y_a - a_J(T))] \\
    &\le 2\left[(\hat Y_a(W)-Y_a(W))^2 + |(\hat Y_a(W) -  Y_a(W))(Y_a(W) - a_J(T))|\right. \\ &\hspace{4em}\left. (\hat a_J(T) - a_J(T))^2 + |(Y_a(W) - a_J(T))(\hat a_J(T) - a_J(T))|\right].
\end{align*}
Thus, 
\begin{align*}
    \|\E_n(\hat Z\hat Z^T - \tilde Z\tilde Z^T)\| &\le \E_n\|\hat Z\hat Z^T - \tilde Z\tilde Z^T\| \\
    & \le 2\E_n\|\bar\phi_J(T)\bar\phi_J(T)^T\|(\hat Y_a(W)-Y_a(W))^2 \\ & \hspace{1em} + 2\E_n\|\bar\phi_J(T)\bar\phi_J(T)^T\||(\hat Y_a(W) -  Y_a(W))(Y_a(W) - a_J(T))| \\
    & \hspace{1em} + 2\E_n\|\bar\phi_J(T)\bar\phi_J(T)^T\|(\hat a_J(T) - a_J(T))^2 \\
    & \hspace{1em} + 2\E_n\|\bar\phi_J(T)\bar\phi_J(T)^T\||(Y_a(W) - a_J(T))(\hat a_J(T) - a_J(T))|.
\end{align*}
The third and fourth terms are $\lesssim_P (|I_1|^{1/(2+\delta)}+l_J)\sqrt{\dfrac{\xi_J^2\log J}{|I_1|}}$, as established in the proof of Theorem 4.6 in \cite{belloni2015some}. The first term can be controlled by 
\begin{align*} E_n\|\bar\phi_J(T)\bar\phi_J(T)^T\|(\hat Y_a(W)-Y_a(W))^2 &\le \max_{i\in I_1}(\hat Y_a(W_i) - Y_a(W_i))^2 \E_n \|\phi_J\phi_J(T)^T\|
\\ &\lesssim_P \|\hat h - h\|^2_{\infty|\hat h} O_p(1) \lesssim_P \|\hat h - h\|^2_{\infty|\hat h}.
\end{align*}
For the second term,
\begin{align*}
    & \E_n\|\bar\phi_J(T)\bar\phi_J(T)^T\||(\hat Y_a(W) -  Y_a(W))(Y_a(W) - a_J(T))| \\ &\le \max_{i\in I_1}|(\hat Y_a(W_i) -  Y_a(W_i))(Y_a(W_i) - a_J(T_i))| \E_n \|\bar\phi_J(T)\bar\phi_J(T)^T\| \\
    &\le \max_{i\in I_1}|\hat Y_a(W_i) -  Y_a(W_i)| \max_{j\in I_1} |Y_a(W_i) -a_J(T_i)| O_p(1) \lesssim_P \max_{i\in |I_1|}|\hat h(W_i)-h(W_i)| (|I_1|^{\frac{1}{2+\delta}}+ l_J) \\ & = \lambda_h (|I_1|^{\frac{1}{2+\delta}}+ l_J).
\end{align*}
Combining all the terms, we obtain 
$$\|\E_n \hat Z\hat Z^T - \E ZZ^T\| \lesssim_P (|I_1|^{\frac{1}{2+\delta}}+l_J)\left(\sqrt{\dfrac{\xi_J^2\log J}{|I_1|}}+\lambda_h\right) + \|\hat h - h\|^2_{\infty|\hat h}.$$
The proof is thus complete by noting that $\|\hat Q-Q\| \lesssim_P\sqrt{\dfrac{\xi_J^2\log\xi_J}{|I_1|}}$, and the definitions of $\Omega$ and $\hat\Omega$.\hfill $\blacksquare$


\end{document}